\documentclass[aps,twocolumn,floatfix,superscriptaddress]{revtex4}
\usepackage{amssymb,amsmath,graphicx,times}
\usepackage{mathptmx}

\usepackage{amsfonts,amsthm} 
\usepackage{bm,bbm}
\usepackage{physics}
\usepackage{mathtools}
\usepackage{lscape}
\usepackage{multirow}
\usepackage{diagbox}
\usepackage{color}
\usepackage{fdsymbol}
\usepackage{setspace}
\usepackage[dvipsnames]{xcolor}
\usepackage{hyperref}
\usepackage{tensor}

\usepackage{natbib}

\usepackage{pifont}
\def\club{\ding{168}}
\def\diamond{\color{red}\ding{169}}
\def\heart{\color{red}\ding{170}}
\def\spade{\ding{171}}
\def\flower{\color{blue}\ding{95}}
\def\star{\color{blue}\ding{87}}

\usepackage{url}
\makeatletter
\g@addto@macro{\UrlBreaks}{\UrlOrds}
\makeatother

\newcounter{mnotecount}[section]
\renewcommand{\themnotecount}{\arabic{mnotecount}}
\newcommand{\mnote}[1]
{\protect{\stepcounter{mnotecount}}$^{\mbox{\footnotesize
$
\bullet$\themnotecount}}$ \marginpar{
\raggedright\tiny\em
$\!\!\!\!\!\!\,\bullet$\themnotecount: #1} }

\definecolor{dg}{rgb}{0,.5,0} 

\usepackage{soul}

\DeclareMathOperator{\PT}{\Gamma}
\DeclareMathOperator{\R}{R}

\def\ket#1{{|#1\rangle}}
\def\bra#1{{\langle#1|}}

\newcommand{\blu}{\color{blue}}

\newcommand{\beq}{\begin{equation}}
\newcommand{\eeq}{\end{equation}}

\usepackage{tikz}

\begin{document}
\title{Thirty-six entangled officers of Euler: Quantum solution to a classically impossible problem}

\author{Suhail Ahmad Rather}
\thanks{Both authors contributed equally}
\affiliation{Department of Physics, Indian Institute of Technology Madras, Chennai 600036, India}
        
\author{Adam Burchardt}
\thanks{Both authors contributed equally}
\affiliation{Institute of Theoretical Physics, 
Jagiellonian University, ul. {\L}ojasiewicza 11, 30--348 Krak\'ow, Poland}

\author{Wojciech Bruzda} 
\affiliation{Institute of Theoretical Physics, 
Jagiellonian University, ul. {\L}ojasiewicza 11, 30--348 Krak\'ow, Poland}

\author{Grzegorz~Rajchel-Mieldzio{\'c}}
\affiliation{Center for Theoretical Physics, Polish Academy of Sciences, Al. Lotnik\'{o}w 32/46, 02-668 Warszawa, Poland}

\author{Arul Lakshminarayan}
\affiliation{Department of Physics, Indian Institute of Technology Madras, Chennai 600036, India}
 
\author{Karol {\.Z}yczkowski}
\affiliation{Institute of Theoretical Physics, 
Jagiellonian University, ul. {\L}ojasiewicza 11, 30--348 Krak\'ow, Poland}
\affiliation{Center for Theoretical Physics, Polish Academy of Sciences, Al. Lotnik\'{o}w 32/46, 02-668 Warszawa, Poland}


\date{August 6, 2021}

\begin{abstract}
The negative solution to the famous problem of $36$ officers of Euler implies that there are no two orthogonal Latin squares of order six. We show that the problem has a solution, provided the officers are entangled, and construct orthogonal quantum Latin squares of this size. As a consequence, we find an example of the long elusive Absolutely Maximally Entangled state AME$(4,6)$ of four subsystems with six levels each, equivalently a $2$-unitary matrix of size $36$, which maximizes the entangling power among all bipartite unitary gates of this dimension, or a perfect tensor with four indices, each running from one to six. This special state deserves the appellation {\sl golden AME state} as the golden ratio appears prominently in its elements. This result allows us to construct a pure nonadditive quhex quantum error detection code $(\!(3,6,2)\!)_6$, which saturates the Singleton bound and allows one to encode a $6$-level state into a triplet of such states. 
\end{abstract}



\maketitle



\noindent {\em Introduction}: 
It is well-known that quantum entanglement leads to peculiar consequences, and enable technical tasks such as
teleportation, secret sharing, secure key cryptography, and error correcting
codes \cite{nielsen_chuang_2010}. Special entangled states such as the cluster states are a resource
for measurement based quantum computing \cite{RaussBriegel_2001,RaussBriegel_2003}.  Multipartite entanglement is also implicated in quantum many-body states and leads to thermalization in isolated systems \cite{Amico_Osterloh_2008,Kaufman_Greiner_2016}. While the understanding of multipartite entanglement is still evolving,
an extreme class of states: the absolutely maximally entangled (AME) ones are clearly singled out and easily defined. AME states find applications in several quantum protocols including 
 quantum secret sharing and parallel teleportation \cite{HelwigAME}, holographic quantum error correcting codes \cite{Pastawski2015HolographicQE,MazurekGrudka}, and quantum repeaters \cite{AlsinaStab}. However, 
they are not easy to construct and often their very existence is unknown.  

A pure quantum state, $\ket{\psi} \in \mathcal{H}_d^{\otimes N}$ 
of $N$ parties, each of a local dimension $d$,
 is {\sl absolutely maximally entangled} \cite{Scott04,HelwigAME},
 written as AME$(N,d)$,
  if it is maximally entangled for every bipartition.
For $N=2$ and $N=3$ parties, generalizations of the Bell and GHZ states, $\sum_{k=1}^d |kk\rangle/\sqrt{d}$ and $\sum_{k=1}^d |kkk\rangle/\sqrt{d}$
are AME states. The smallest number of parties of interest is therefore $N=4$ where it is known that AME$(4,2)$ does not exist, four qubits cannot be
absolutely maximally entangled \cite{HIGUCHI2000,Huber_2018}. However for all $d>2$, except for $d=6$, AME states of 4 parties are known to exist. Thus arguably the most interesting open case was the existence of AME$(4,6)$. It has featured on open-problem lists of quantum information \cite{AME_IQOQI,horodecki_2020_open} and in this Letter we settle this
positively by explicitly constructing an example, leaving only four qubits as exceptional.

To shed light on the peculiar properties of such states, consider a collection of four dice.
Alice selects {\em any} two dice and rolls them, obtaining one of 36 equally likely outcomes, as Bob rolls the remaining ones. If the entire state is AME$(4,6)$,
 Alice can always deduce the result obtained in Bob's part of the $4$-party system.


\begin{figure}[htbp]
        \centerline{ \hbox{
                 \includegraphics[scale=.72]{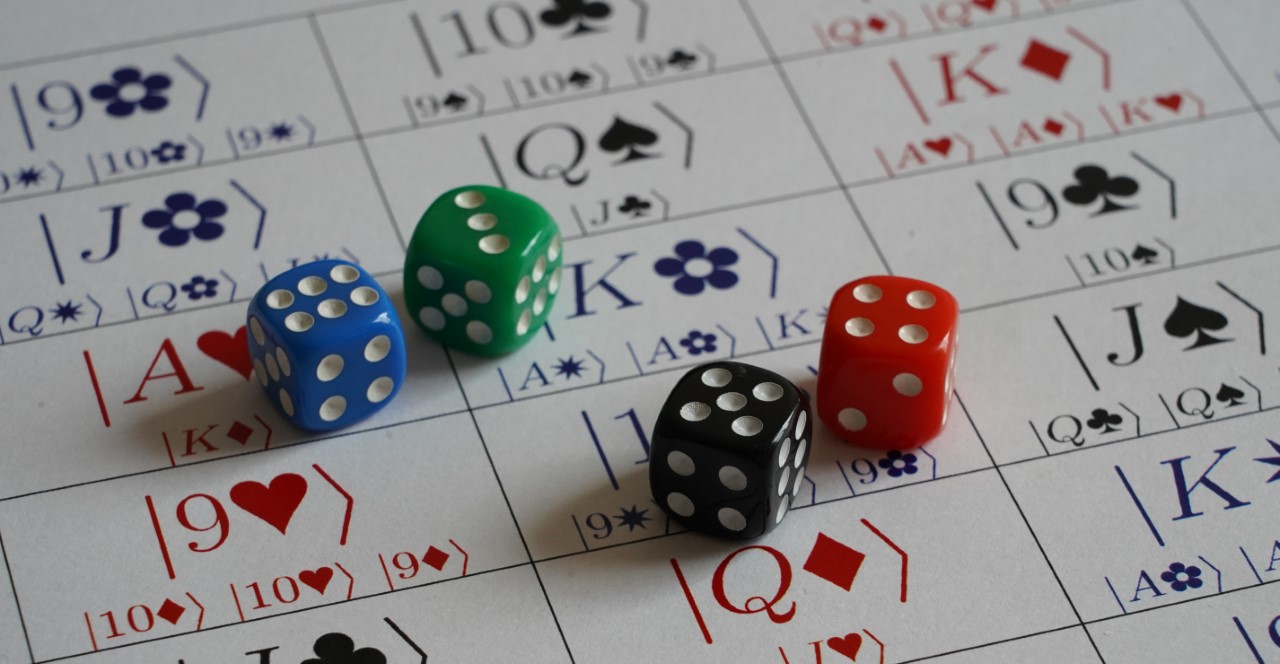}
                   }}
        \caption{Four dice in the golden absolutely maximally entangled state of four-quhex, AME$(4,6)$, corresponding to $36$ entangled
officers of Euler. Any pair of dice is  unbiased, although their outcome determines the state of the other two. This is not possible if the four dice are replaced by four coins (qubits).}
        \label{fig:dice}
\end{figure}
\medskip

Furthermore, such a state allows one 
to teleport any unknown, two-dice quantum state,
 from any two owners of two subsystems 
 to the lab possessing the two other 
 dice of the entangled state of the four-party system. These are not possible
 if the dice is replaced by two-sided coins as AME$(4,2)$ states do not exist
 \cite{HIGUCHI2000}.

The existence of AME$(4,d)$ for all $d>2$ and $d\neq 6$ has been
known from the existence of particular combinatorial designs, namely orthogonal 
Latin squares (OLS), also known as Graeco-Latin squares. 
A simple example of a combinatorial design \cite{CD07} is given by a single {\sl Latin square}:
a $d\times d$ array   
filled with $d$ copies of $d$ different symbols, such that 
each occurs once in each row and each column.
For the design of experiments, one uses Graeco-Latin squares: 
two Latin squares arranged in such a way
that the ordered pairs of entries in all cells of the square are distinct.
The name of the design refers to a popular way to
represent such a pair by one Greek character and one Latin.

\begin{figure}[h!]
	\centering
\begin{equation} \
\large
\begin{array}{|c|c|c|} \hline 
{\beta\; C}    & \gamma \; A &{ \alpha \; B} \\ \hline
{ \gamma\; B} & \alpha\; C & { \beta \; A}\\ \hline
{ \alpha\; A}  & \beta \; B& {\gamma \; C}\\ \hline
\end{array} 
\ = \ 
\begin{array}{|c|c|c|} \hline 
{ \color{red} K \text{\diamond} }    & Q \text{\spade} &  A \text{\club}   \\ \hline
{ Q\text{\club}}  & {\color{red} A \text{\diamond}} &  K \text{\spade}    \\  \hline
{ A \text{\spade}}  & K \text{\club} &  {\color{red}  Q \text{\diamond} }  \\  \hline
\end{array} 
\ = \ 
\begin{array}{|c|c|c|} \hline 
2, {\color{blue}3}   & 3,{\color{blue}1}&  1, {\color{blue} 2}  \\ \hline
3, {\color{blue}2}  & 1, {\color{blue}3} &  2,{\color{blue} 1} \\  \hline
 1,{\color{blue}1}  & 2, {\color{blue}2}&  3,{\color{blue}3}  \\  \hline
\end{array} 
\nonumber
 \end{equation}
\caption{An example of Graeco-Latin square of order $d = 3$,
In the middle, Greek and Latin letters are replaced by 
ranks and suits of cards, on the right by pairs of numbers.}
\label{fig:OLS3}
\end{figure}
It is easy to see that there are no Graeco-Latin squares 
 of size two.
Furthermore, it is not difficult to construct such a combinatorial design
 for $d=3$ -- see  Fig.~\ref{fig:OLS3}.
Analogous constructions also work for $d=4,5$,
and in general, for odd numbers $d$ and multiples of four \cite{DK91}.
The case of $d=6$ is thus special.
 Leonhard Euler examined 
the now-famous problem \cite{Euler36}: 
``Six different regiments have six officers, each one belonging to different ranks. 
Can these 36 officers be arranged in a square formation so that each row and column contains one officer of each rank and one of each regiment?'' 
As Euler observed, such an arrangement, equivalent to a {Graeco-Latin square} of order $6$, written OLS(6), does not exist, which was proven much later by Gaston Tarry \cite{GastonTarry}. 
 Two OLS of size $d=10$ were constructed in the 20th century
 and it is known that such designs 
 exist  \cite{JCD01} for any natural number $d\neq 2$, or $\neq 6$. This enables the construction of AME$(4,d)$ for any $d$ other than $d=2$ and $d=6$ \cite{Clarisse_2005}.
 

While classical designs are built of discrete objects, their quantum analogs, introduced in the seminal thesis of Zauner \cite{Za99},
are composed of pure quantum states: normalized 
vectors from a complex Hilbert space  $\mathcal{H}_d$ of dimension $d$.
Such distinguished configurations of states play a significant role
in quantum information, as they describe generalized quantum measurements
with special properties.
In particular,  quantum analogs of Latin squares 
were introduced in \cite {MV16}, while various approaches to 
 orthogonal quantum  LS were advocated in \cite{GRMZ18,MV19,Ri20}.

In this work, we formulate a quantum analog of Euler's problem of 36 officers 
\cite{GRMZ18} and present its complete analytic solution. 
As AME$(4,2)$ does not exist and standard OLS exist for any natural $d$,
different from $2$ and $6$, the latter case corresponding to the
quantum version of the Euler problem of 36 officers,
was the only open problem for  $4$-partite systems \cite{Table_AME,yu2020complete}. 
Our construction of an AME$(4,6)$ state, shows that for 4 parties,
$d=6$ is the only dimension in which there are
no classical OLS, but there exists a quantum one. 
Even though OLS of order six do not exist, we present a coarse-grained OLS of that order, whose structure stands behind its constructed quantum analog.  

\smallskip
\noindent{\em Multipartite entangled states from Graeco-Latin squares}: A pure quantum state
$|\psi\rangle \in \mathcal{H}^A_d \otimes \mathcal{H}^B_d$
 of a bipartite  $d \times d$ system can be expanded in a product basis,
$|\psi\rangle =\sum_{i,j=1}^d C_{ij} |i \rangle |j\rangle$,
where the matrix $C$ of coefficients satisfies the normalization
condition, $\|C\|^2={\rm Tr}\, CC^{\dagger}=1$. A pure state is maximally entangled
if its partial trace is maximally mixed,
so that the matrix of coefficients is unitary up to rescaling,
 $CC^{\dagger}= I/d$.
A pure quantum state, $\ket{\psi} \in \mathcal{H}_d^{\otimes N}$ 
of $N$ parties, each of a local dimension $d$ is AME$(N,d)$
if the partial trace $\text{Tr}_S \ket{\psi}\bra{\psi} \propto I$, for any subsystem $S$ of $|S|=\lfloor N/2 \rfloor$ parties. 

Any Graeco-Latin square of order $d$ determines a
$4$-party  quantum state $\ket{\Psi} \in \mathcal{H}_d^{\otimes 4}$,
\begin{equation}
\label{state}
\ket{\Psi}=
\frac{1}{d} 
\sum_{i,j,k,\ell =1}^{d}
T_{ijk \ell}
\ket{i}\ket{j} \ket{k}\ket{\ell} ,
\end{equation}
where coefficients
$T_{ijk \ell}=1$
if the pair $(k,\ell)$ is an entry in $i$-th row and $j$-th column, while
$T_{ijk \ell}=0$
otherwise. 
Note that the matrix of order $d^2$,
 indexed by doubled indices, 
$U_{ij, \: k \ell}:=T_{ijk \ell}$,
forms a permutation. 
Orthogonality conditions imposed on Latin squares
implies that the matrices corresponding to the other two bipartitions of the four indices into pairs:
 $ik|j \ell$ and $i \ell|jk$, also form permutations.
 This follows as every relation between any two pairs of features (such as the column number and the card suit versus 
 the row number and the card rank shown in Fig.~\ref{fig:OLS3}) is a bijection.
 
It is convenient to recall reorderings of entries of a matrix of size $d^2$
 used in quantum theory \cite{BZ17}.
Representing the matrix in a product basis as 
$U_{ij,k \ell}=\langle i,j|U| k, \ell  \rangle$,
one defines the  partially transposed matrix, $U^{\Gamma}$
 and the reshuffled matrix $U^{\R}$,
\begin{equation}
\label{Gamma}
U^{\Gamma}_{ij,k \ell}=U_{i\ell, k j}, \ \ \  \ 
U^{\R}_{ij,k \ell}=U_{ik,j \ell}.
\end{equation}
Making use of this notation, one can say that a permutation matrix $U$
of order $d^2$ yields an OLS$(d)$ if the reordered matrices,
 $U^{\R}$ and $U^{\Gamma}$ are also permutations.
In this case, 
the $4$-partite state of Eq.~\ref{state} has a particular property:
It is maximally entangled with respect to any bipartition 
of four indices $i,j,k,\ell$ into two pairs, and is hence an AME$(4,d)$ state.

\smallskip
\noindent{\em Orthogonal quantum Latin squares}:
To obtain other AME states of this class, 
 we retain the condition that the matrices $U,U^{\R}$ and $U^\Gamma$ be unitary while relaxing the condition that they are permutations. 
Such a matrix $U$  is called 2-unitary \cite{MultiUnitary}, see  supplemental material section~I for more details.
 In fact, any $2$-unitary matrix $U\in \mathbb{U}(d^2)$ yields an
AME$(4,d)$ state 
 \begin{equation}
\label{state22}
|\text{AME}(4,d)\rangle=
\frac{1}{d} 
\sum_{i,j=1}^{d}
\ket{i}\ket{j} \ket{\psi_{ij}} ,
\end{equation}
where $\ket{\psi_{ij}}= U \ket{i}\ket{j}$. 
 This allows one to say that 
 the corresponding quantum design of $d^2$ bipartite quantum states
 $|\psi_{ij}\rangle$, $i,j=1,\dots, d$, forms an orthogonal quantum  Latin square 
(OQLS) 
  -- for a  formal definition see supplemental material (SM) section~I.

It may be noted that a 2-unitary matrix $U$ of size $d^2$, treated as a $4$-index tensor, $T_{ijk\ell}:=U_{(ij), (k\ell)}$ has been called a {\sl perfect tensor} \cite{Pastawski2015HolographicQE} and used in constructing quantum error correcting codes. Any of its $2$-index flattenings
 $T_{ij}^{k\ell}, T_{ik}^{j\ell}, T_{i\ell}^{jk}$ is unitary and provides an \textit{isometry} between any pair of its indices.
The partial trace of the $4$-party state $\ket{\Psi}$ related to a perfect tensor
as in Eq.~\ref{state22} is maximally mixed for any symmetric bipartition of the system.
Thus existence of such a perfect tensor with $4$ indices running from $1$ to $d$
 is equivalent to the existence of an AME state of four qudits and OQLS of size $d$.

\noindent{\em Searching for 36 entangled officers}:
To tackle the quantum Euler problem of AME$(4,6)$, equivalent to finding a $2$-unitary matrix of order $36$,
we used an iterative numerical technique based on nonlinear maps in the space of unitary matrices $\mathbb{U}(d^2)$ introduced recently in \cite{SAA2020}.
A closely related Sinkhorn-like algorithm to generate unitary matrices such that only their partial transpose is unitary
was presented earlier in \cite{BNechita}.
One map that produces 2-unitaries $\mathcal{M}_{\PT \R}: U_0  \mapsto U_1$, 
consists of two parts (i) $\R$ and ${\PT}$ operations in that order $U_0 \mapsto U_0^{\R} \mapsto \left(U_0 ^{\R} \right)^{\PT} :=U_0^{\PT \R}$, and (ii) projection onto the nearest unitary matrix $U_0^{ \PT \R} \mapsto U_1$ using the polar decomposition, $U_0^{\PT \R}=U_1 H$, where $H$ is a positive semi-definite matrix. It is straightforward to see that 2-unitary matrices are fixed points of the map $\mathcal{M}_{\PT \R}^3$, or period-3 orbits of $\mathcal{M}_{\PT \R}$. After $n$ iterations of the map, $\mathcal{M}_{ \PT \R}^n\left[U_0\right]=U_n$, and as $n \rightarrow \infty$ this converges to 2-unitary matrices with high probabilities for $d=3$ and $d=4$, even using seeds $U_0$ sampled randomly according to the Haar measure on the unitary group \cite{SAA2020}.

The key to generating a $2$-unitary matrix for $d>4$ is to choose an appropriate
seed matrix. The neighborhood of permutation matrices approximating 
OLS is a natural choice. In the most interesting case of $d=6$, for which
there are no OLS, seeds in the vicinity of the permutation matrix $P_{36}$ \cite{Clarisse_2005} that is closest to an OLS, surprisingly, do not lead to a 2-unitary. However, there do exist other suitable permutation matrices whose vicinity contains seeds that  
 under the map $\mathcal{M}_{\PT \R}$  converge to $2$-unitaries.
An example of a seed that leads to the $2$-unitary solution displayed in this work is provided in section III of SM.

A 2-unitary matrix remains 2-unitary on multiplication by local unitary operators. 
Using this freedom, we applied a search algorithm over the group 
$\mathbb{U}(6) \otimes \mathbb{U}(6)$ of local unitary operations, 
to orthogonalize certain rows and columns in a given numerical 2-unitary matrix $U$ and its rearrangements $U^{\R}$, $(U^{\R})^{\PT}$. 
We searched for block structures in all these three matrices abstracting from the exact form of the matrices. 
The particular choice of the orthogonality relations corresponds to the block structure of the eventually obtained analytical solution. 
These tools can be generalized to construct multi-unitary operators and corresponding AME states in other local dimensions and number of parties. While the solution presented is
the smallest one of arguably greatest interest, namely AME$(4,6)$, the methods can potentially yield maximally entangled states that are not 
created by presently known techniques. 

\begin{figure*}
\centering
\includegraphics[scale=.95]{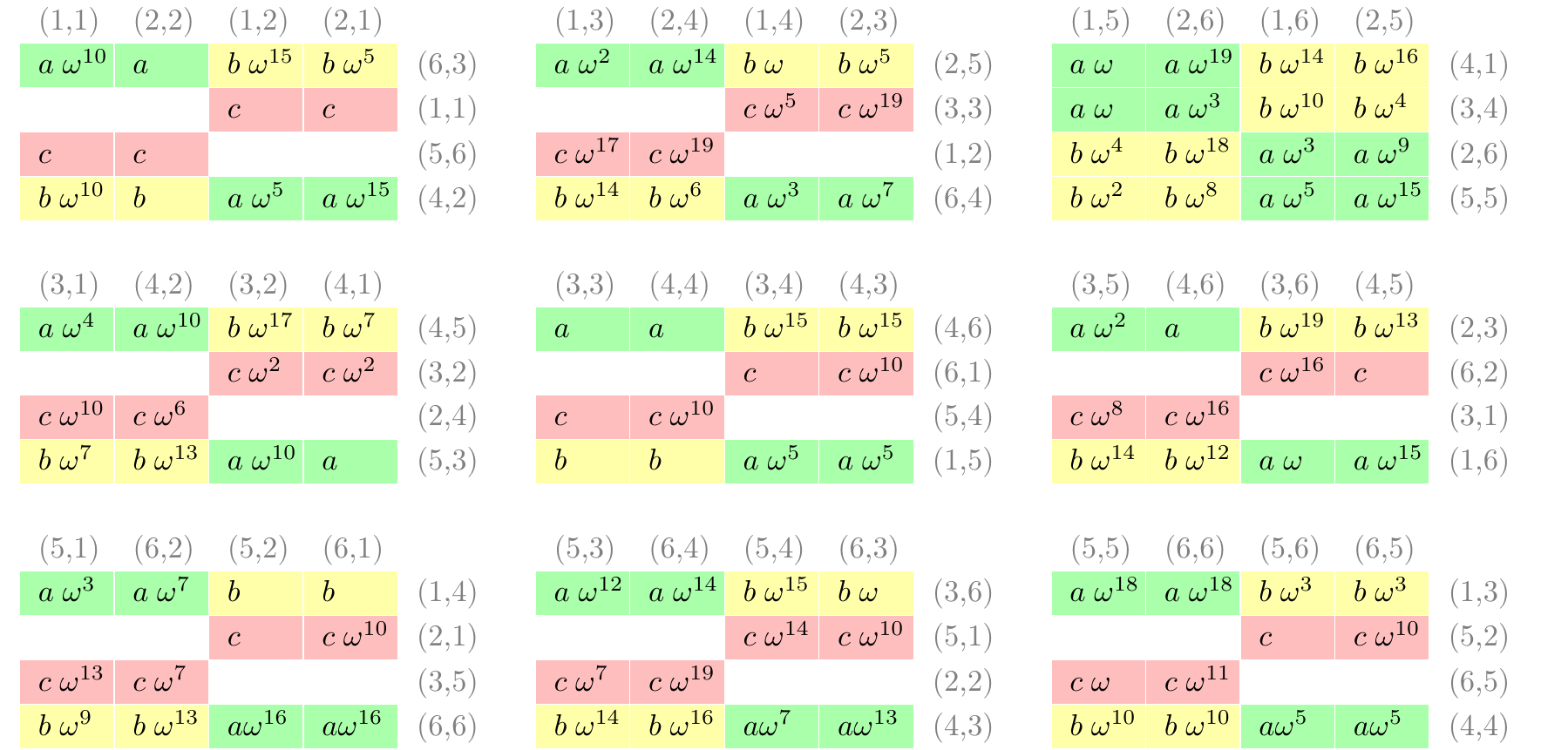}
\caption{Non-vanishing coefficients $T_{ijk \ell}$ 
of the AME$(4,6)$ state $\ket{\Psi}$. Indices $(i,j) $ are indicated in rows, $(k,l) $ in columns, respectively. 
Treating each coordinates $(i,j)$ as a position (row and column) of Euler's officer, its rank and regiment are in superposition of two or four canonical ranks and regiments. Equivalently, the picture shows non-zero entries of $2$-unitary matrix $\mathcal{U}^{}_{36}$ of size $36$ where $j+6(i-1)$ labels the relevant row while $\ell+6(k-1)$ determines the column. 
Note that all $36$ officers, each represented by a single row of the matrix, are maximally entangled qubit states, $16$ of them have minimal support being bipartite Bell states (red rows), while the remaining $20$ have maximal support (yellow/green rows) -- see Eq.~\ref{eq:QOLS6} and SM.
The order of the blocks reflects an additional structure as
  the numbering of columns increases   by six as we proceed to 
  a block below.
 Depicted matrix preserves its structure of nine unitary blocks 
of size four  with respect to transformations of
 reshuffling $j\longleftrightarrow k$, and partial transpose $j \longleftrightarrow \ell$. 
}
\label{fig2}
\end{figure*}

\noindent{\em Solution found}: To present the solution to the problem of the 36 entangled officers of Euler, we display the 
coefficients $T_{ijk \ell}$ of the AME$(4,6)$ quantum state $\in \mathcal{H}_6^{\otimes 4}$ (four quhex state), see Eq.~\ref{state22}. 
Non-vanishing coefficients 
$T_{ijk \ell}$ might be conveniently written in form of a table, see Fig.~\ref{fig2}. 
The provided construction is based on the root of unity of order 20, denoted by $\omega=\exp(i \pi/10)$. 
There exist only three non-zero amplitudes:
\begin{align}
\label{abc}
a=&\left(\sqrt{2}(\omega+\overline{\omega})\right)^{-1}=\left(5+\sqrt{5}\right)^{-1/2}, \nonumber \\
b=&\left(\sqrt{2}(\omega^3+\overline{\omega}^{3})\right)^{-1}=\left((5+\sqrt{5})/20\right)^{1/2},   \\
c=&1/\sqrt{2},\nonumber
\end{align}
whose numerical values are  $ \simeq (0.3717, \, 0.6015, \,0.7071)$, respectively. The bar over the symbol indicates the complex conjugate.
The relations $a^2+b^2= c^2=1/2$  
 and $b/a=\varphi=(1+\sqrt{5})/2$ the golden ratio, determine  all amplitudes appearing in the solution and explain why the constructed AME state deserves to be called the {\sl golden} AME {\sl state}.
Checking the property of being an AME state reduces to verification of several equations involving roots of unity of order 20, which we elaborate in a detailed way in 
SM section~II. 
The key result of this work consists, therefore, of the following
assertion proved in SM: 

\smallskip
{\bf Theorem 1}. 
{\sl There exists an} \text{AME}$(4,6)$ {\sl state of $4$ parties with $6$ levels each.} 

\smallskip
\noindent
This statement is equivalent to the existence of a $2$-unitary matrix $\mathcal{U}_{36} \in \mathbb{U}(36)$,
and a perfect tensor $T_{ijkl} := \mathcal{U}_{36\,(ij)(kl)}$,
and a solution to the quantum analog of the 36-officers problem of Euler.

An explicit solution of the generalized Euler problem can be conveniently written in terms of a \textit{quantum} OLS in the form of 36 states $\ket{\psi_{ij}}=\mathcal{U}_{36}|i\rangle |j \rangle \in \mathcal{H}_6^{\otimes 2}$ representing the entangled officers. Following Euler's notation used for Graeco-Latin squares we label the rank of each officer by a rank of the card A, K, Q, J, 10, 9, and (extended set of) suits {\spade}, {\club}, {\diamond}, {\heart}, {\flower}, {\star}
are used to label her regiment: 
{\small 
\begin{eqnarray}
\label{eq:QOLS6}
 |\psi_{11}\rangle &=& c \ket{\text{A\club}} + c \ket{\text{K\spade}}, \nonumber \\
 |\psi_{12}\rangle  &=& c \omega^{17} \ket{{\color{red}\text{A\diamond}}} + c \omega^{19} \ket{{\color{red}\text{K\heart}}},    \\
  & \vdots&  
  \nonumber \\
|\psi_{66}\rangle  &=&  
  b\omega^9 \ket{\text{10 \spade}}+
  b\omega^{13} \ket{\text{9 \club}} + 
  a \omega^{16} \ket{\text{10 \club}} + 
  a \omega^{16} \ket{\text{9 \spade}}, \nonumber
\end{eqnarray}
}
\noindent
while expressions for the remaining thirty-three states can be directly read out from the $2$-unitary matrix presented in
Fig.~\ref{fig2}. 
Notice that the related state in Eq.~\ref{state22} is an AME state. 
{Remarkably, all the 36 states $|\psi_{ij}\rangle$ are maximally 
entangled as two-qubit states. While this is evident for states such as $|\psi_{11}\rangle$, even $|\psi_{66}\rangle$ and others
with support on 4 states are maximally entangled, thanks to the special value of the phases.}

\begin{figure*}
\centering
\includegraphics[scale=0.85]{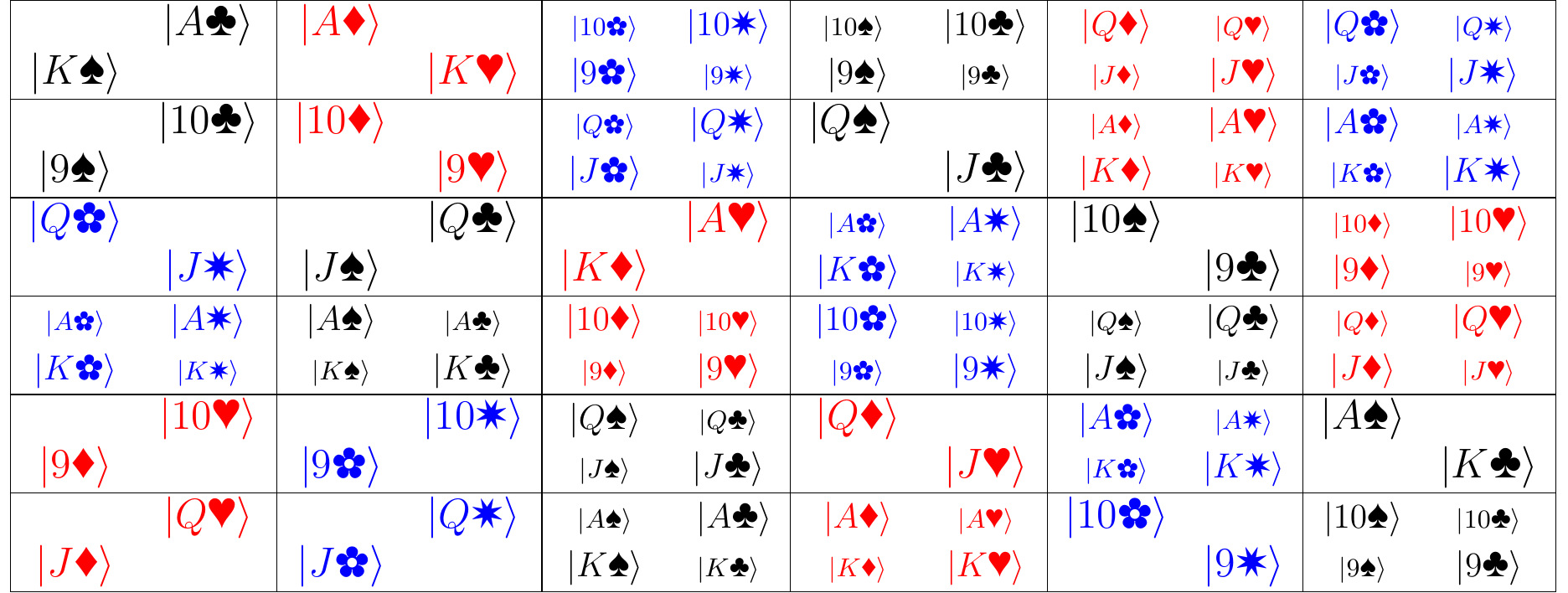}
	\caption{Visualization of entanglement between Euler's officers. 
Position of each officer: $i$-th row and $j$-th column, is presented in appropriate row and column on the array. 
Ranks and regiments of officers (relevant to the indices $k$ and $\ell$ in $T_{ijk \ell}$) are represented in form of cards with corresponding ranks and suits  
(in particular, 
$1\rightarrow A / \text{\spade} $, 
$2\rightarrow K / \text{\club}$, 
$3\rightarrow Q / {\color{red} \text{\diamond}}$, 
$4\rightarrow J / {\color{red} \text{\heart} }$,
$5\rightarrow 10 / {\color{blue} \text{\flower} }$, 
$6\rightarrow 9 / {\color{blue}  \text{\star}}$, 
for $k/\ell$ respectively). 
The grade of each officer is in a superposition of four basic grades: two ranks, and two regiments. 
The font size of each ket corresponds to the amplitude of the related tensor element. 
 Notice that a classical solution to Euler's problem would correspond to the array with only one card in each entry.
	Note the structures arising in the array, e.g. aces are entangled only with kings, queens with jacks, and 10s with 9s.
	Moreover, the \emph{colors} of the cards are not entangled with each other. 
	Hence, officers are grouped into nine sets with four elements, each sharing the same pairs of figures and colors of the card. 
	For example, officers on positions $(1,1),(4,2),(5,6),(6,3)$ share two black figures $ A,K$ of a black suits $ \text{\spade} ,\text{\club}$, which corresponds to the top-left clique on Fig.~\ref{fig2}.
	}
\label{fig4}
\end{figure*}

\noindent{\em Structure of the AME$(4,6)$ state}: Recall that a classical OLS corresponds to a $2$-unitary permutation matrix. 
Since there is no solution to the original problem of Euler, the $2$-unitary permutation matrix of size 36 does not exist. 
Nonetheless, we can present the AME$(4,6)$ state obtained by us in a form similar to the classical solution of AME$(4,3)$ generated from classical OLS in Fig.~\ref{fig:OLS3}.
Let us consider every row of a 2-unitary matrix as a place to put an ``officer" in, then we can express the entanglement in our solution by showing which two (or four) officers are entangled, thus producing Fig.~\ref{fig4}.

The $2$-unitary matrix $\mathcal{U}_{36}$ described here has, up to permutations, 
the structure of nine unitary blocks 
of size $4$. 
Moreover, the block structure is also characteristic
for the reshuffled matrix $\mathcal{U}^{\R}_{36}$ and the partially 
transposed matrix $\mathcal{U}^{\Gamma}_{36}$. In other words, 
 in the original matrix $\mathcal{U}_{36}$ we 
  found the block structure \textit{invariant} 
under reshuffling and partial transpose. 
The problem of finding 36 entangled officers of Euler splits into two sub-problems: to identify first a block-invariant structure
and then to select adequate non-zero elements within them.
A particular combinatorial design underlies the invariant structure.
Grouping symbols of indices $k$ and $\ell$ in the presented perfect tensor $T_{ijk \ell}$ in pairs: $1,2\rightarrow A /\alpha$, and $3,4\rightarrow B / \beta$, and $5,6\rightarrow C/ \gamma$ for $k/ \ell $   respectively, results in a coarse-grained OLS, which reveals the described block structure, see Fig.~ \ref{fig3}.  

\begin{figure}[h!]
	\centering
\begin{equation} \
\large
\begin{array}{|c|c|c|c|c|c|} \hline 
{ A\alpha}  & A\beta& {C\gamma}&
{ C\alpha}  & B\beta& {B\gamma}\\ \hline
{ C\alpha}  & C\beta& {B\gamma}&
{ B\alpha}  & A\beta& {A\gamma}\\ \hline
{ B\gamma} & B\alpha  & { A\beta}&
{ A\gamma} & C\alpha  & { C\beta}\\ \hline
{ A\gamma} & A\alpha  & { C\beta}&
{ C\gamma} & B\alpha  & { B\beta}\\ \hline
{C\beta}    & C\gamma  &{B \alpha} &
{B\beta}    & A\gamma  &{A \alpha} \\ \hline
{B\beta}    & B\gamma  &{A \alpha} &
{A\beta}    & C\gamma  &{C \alpha} \\ \hline
\end{array} 
\nonumber
 \end{equation}
\caption{A coarse-grained OLS of order 6, which reveals the block structure of a perfect tensor $T_{ijk \ell}$. 
Indices of non-vanishing elements of the tensor $T_{ijk \ell}$ are presented: $i$ in row, $j$ in column, while a pair of coarse-grained indices $k,\ell$ in the entry. 
Each pair of symbols repeats exactly four times in the array. 
Moreover, each symbol on each position repeats exactly twice in each row and column. 
Notice that the array above corresponds to Fig.~\ref{fig4} by coarse-graining figures 
$A/K\rightarrow A; \;
Q/J\rightarrow B; \;
10/9\rightarrow C$,
and suits 
$\text{\spade}/\text{\club} \rightarrow \alpha; \;
{\color{red} \text{\diamond}} / {\color{red} \text{\heart} }\rightarrow \beta; \;
{\color{blue} \text{\flower} } /{\color{blue}  \text{\star}}\rightarrow \gamma $ 
of cards in entries.
}
\label{fig3}
\end{figure} 

\smallskip
\noindent{\em Quantum codes}: Several examples of quantum error correction codes  
discussed in the literature belong to the class of
 \textit{additive (stabilizer) codes}, 
quantum analogs of classical additive codes \cite{Cross_2009,AlsinaStab}. 
In particular all hitherto known AME states, are either stabilizer states \cite{AlsinaStab}, or might be derived from the stabilizer construction \cite{BurchardtRaissi20}. 
Stabilizer codes have the structure of an eigenspace of an abelian group generated by multilocal generalized Pauli operators. 
The stabilizer approach is especially effective for codes with local prime power dimensions~$d$.
Stabilizers of an additive code might be presented in their standard form \cite{AlsinaStab}. 
We examined all stabilizer sets of four quhex in their standard form and did not find an AME$(4,6)$ state. 
Therefore, the presented AME$(4,6)$ state is a nonadditive one. 

Nonadditive 
quantum codes are in general more difficult to construct, however, in many cases, they outperform the stabilizer codes \cite{NonadditiveCodes}. 
Thus far, the stabilizer approach practically contained the combinatorial approach to constructing AME and $k$-uniform states.
As we demonstrated, the consideration of coarse-grained combinatorial structures might be successful in constructing genuinely entangled states and have advantages over the stabilizer approach.

In order to successfully use a pure $(\!(4,1,3)\!)_6$ code in an error correction scheme, one may apply the shortening procedure \cite{Rains1999NonbinaryQC,Huber_2020} and obtain $(\!(3,6,2)\!)_6$ code. 
In such way, a single quhex $\ket{i} \in \mathcal{H}_6$ is encoded into a three quhex state $\ket{\tilde i} \in \mathcal{H}_6^{\otimes 3}$
defined by 
\begin{equation*}
\ket{i}  \rightarrow \ket{\tilde i} := \frac{1}{\sqrt{6}} \sum_{j,k,\ell=1}^6 T_{ijk \ell} \ket{j,k,\ell} .
\label{code}
\end{equation*}
Both codes, the initial and shortened ones, are optimal, meeting the quantum Singleton bound \cite{Huber_2020}.
Thus, the presented construction
of AME states of four subsystems with $6$ levels each
 sheds some light on how to construct nonadditive quantum error correction 
 codes, in a case for which the stabilizer approach fails.

\smallskip
\noindent {\em Summary and Outlook}: The famous combinatorial 
problem of 36 officers was posed by Euler, who 
claimed in 1779 that no solution exists. 
The first paper with proof of this statement, by Tarry \cite{GastonTarry}, came only 121 years later, in 1900.
After another 121 years, we have presented a solution to the quantum version wherein the officers can be entangled. 
This unexpected result implies constructive solutions to the related problems of the existence of 
absolutely maximally entangled states of four subsystems with six levels each, a $2$-unitary matrix $\mathcal{U}_{36}$ of size $36$ with maximal entangling power 
and a perfect tensor $T_{ijk \ell}$ with four indices, each running from one to six.
Our results allowed us to construct original quantum error correction codes: a pure code $(\!(4,1,3)\!)_6$, and a shortened code $(\!(3,6,2)\!)_6$, 
which allows encoding a $6$-level state into a set of three such subsystems. It is tempting to believe that the quantum design presented here will trigger further research on quantum combinatorics.

\smallskip
\hspace{3cm}

\acknowledgments{It is a pleasure to thank S.~Aravinda, J.~Czartowski,
D.~Goyeneche,  M.~Grassl, F.~Huber, P.~Mazurek, Z.~Pucha{\l}a and 
A.~Rico for several inspiring discussions and helpful remarks.
Financial support by Narodowe Centrum Nauki 
under the Maestro grant number DEC-2015/18/A/ST2/00274, 
by Foundation for Polish Science 
under the Team-Net project no. POIR.04.04.00-00-17C1/18-00
and by the Department of Science and
Technology, Govt. of India, under grant number
DST/ICPS/QuST/Theme-3/2019/Q69 are gratefully acknowledged.} 


\begin{thebibliography}{99}

\bibitem{nielsen_chuang_2010}
M.~A.~Nielsen, I.~L.~Chuang,
Quantum Computation and Quantum Information, Cambridge University Press (2010).

\bibitem{RaussBriegel_2001}
R.~Raussendorf and H.~J.~Briegel,
A One-Way Quantum Computer
{\em Phys. Rev. Lett.} {\bf 86}, 5188 (2001).

\bibitem{RaussBriegel_2003}
R.~Raussendorf, R.~Browne and H.~J.~Briegel,
Measurement-based quantum computation on cluster states
{\em Phys. Rev.} {\bf A 68}, 022312 (2003).

\bibitem{Amico_Osterloh_2008}
L.~Amico, R.~Fazio, A.~Osterloh, and V.~ Vedral,
Entanglement in many-body systems,
{\em Rev. Mod. Phys.} {\bf 80} 517 (2008).

\bibitem{Kaufman_Greiner_2016}
A.~M.~Kaufman, M.~E.~Tai, A.~Lukin, M.~Rispoli, R.~Schittko, P.~M.~Preiss, and M.~Greiner,
Quantum thermalization through entanglement in an isolated many-body system,
{\em Science} {\bf 353} 6301 (2016).

\bibitem{HelwigAME}
W.~Helwig, W.~Cui, A.~Riera, J.~Latorre, and H.K. Lo, 
 Absolute maximal  entanglement and quantum secret sharing,
   {\em Phys. Rev.} {\bf  A 86},  052335  (2012).

\bibitem{Pastawski2015HolographicQE}
F.~Pastawski, B.~Yoshida, D.~C. Harlow, and J.~Preskill, 
Holographic quantum
  error-correcting codes: toy models for the bulk/boundary correspondence,
  {\em Journal High Energy Phys.}, {\bf 2015}, 1 (2015).

\bibitem{MazurekGrudka}
P.~Mazurek, M.~Farkas, A.~Grudka, M.~Horodecki, and
  M.~Studzi\ifmmode~\acute{n}\else \'{n}\fi{}ski,
  Quantum error-correction codes and absolutely maximally entangled states,
   {\em Phys. Rev.} {\bf A 101}, 042305  (2020).


\bibitem{AlsinaStab} D.~Alsina and M.~Razavi,
Absolutely maximally entangled states,
  quantum-maximum-distance-separable codes, and quantum repeaters,
   {\em Phys. Rev.} {\bf  A 103},  022402 (2021)

\bibitem{Scott04}  A.~J. Scott, 
Multipartite entanglement, quantum-error-correcting codes, and
  entangling power of quantum evolutions,
   {\em Phys. Rev.} {\bf  A 69}, 052330 (2004).

\bibitem{HIGUCHI2000}
A.~Higuchi and A.~Sudbery, 
How entangled can two couples get?, 
{\em Phys.  Lett.} {\bf  A 273},  213  (2000).

\bibitem{Huber_2018}
F.~Huber, C.~Eltschka, J.~Siewert, and   O.~G{\"u}hne,  Bounds on absolutely
  maximally entangled states from shadow inequalities, and the quantum
  {MacWilliams} identity, {\em J. Phys.} {\bf 51}, 175301 (2018). 

\bibitem{AME_IQOQI}
List of {O}pen {Q}uantum {P}roblems, 
 \href{https://oqp.iqoqi.univie.ac.at/existence-of-absolutely-maximally
  -entangled-pure-states}{{P}roblem 35,}
{I}{Q}{O}{Q}{I}  {V}ienna.

\bibitem{horodecki_2020_open}
P.~Horodecki, {\L}.~Rudnicki, and K.~{\.Z}yczkowski,
Five open problems in quantum information,
arXiv:2002.03233. 
  
\bibitem{CD07}  C. J. Colbourn and  J. H. Dinitz (eds.),
{\sl Handbook of Combinatorial Designs},
CRC Press, Boca Raton (2007).

\bibitem{DK91} J. D{\'e}nes and A. D. Keedwell (eds.), 
{\sl Latin Squares: New
Developments in the Theory and Applications}, 
North-Holland, Amsterdam, (1991).

\bibitem{Euler36}
L.~Euler, ``Recherches sur une nouvelle espece de quarres magiques,'' {\em
  Verhandelingen uitgegeven door het zeeuwsch Genootschap der Wetenschappen te
  Vlissingen 9}, Middelburg, 85–239 (1782).
\newblock Also published in {\em Commentationes Arithmeticae} 2, 302–361 1(849).
  Available online in The Euler Archive: \url{http://
  eulerarchive.maa.org/pages/E530.htm}l. Accessed, April 2021.

\bibitem{GastonTarry}
G.~Tarry, ``Le probl{\'e}me de 36 officiers,'' {\em Compte Rendu de l'Association
  Française pour l'Avancement des Sciences}
  . Secr{\'e}tariat de l'Association.
   {\bf 1}, 122 (1900).

\bibitem{JCD01}
C.~{J. Colbourn} and J.~{H. Dinitz}, 
 Mutually orthogonal {L}atin squares: a brief survey of constructions, 
  {\em J.  Stat. Planning   Inference} {\bf  95},  9 (2001).

\bibitem{Clarisse_2005}
L.~Clarisse, S.~Ghosh, S.~Severini, and A.~Sudbery,
Entangling power of  permutations,
 {\em Phys. Rev.}, {\bf A 72} (2005).

\bibitem{Za99} G. Zauner, Quantendesigns: Grundz\"uge einer
nichtkommutativen Designtheorie, Ph. D. thesis, Universit\"at Wien (1999).

\bibitem{MV16} B. Musto and J.Vicary, 
Quantum Latin squares and unitary error bases,
{\sl Quantum Inf. Comput.} {\bf 16}, 1318 (2016).

\bibitem{GRMZ18} D.~Goyeneche, Z.~Raissi, S.~Di~Martino, 
and K.~{\.Z}yczkowski, 
Entanglement and quantum combinatorial designs,
 {\em  Phys. Rev.} {\bf  A 97},  062326 (2018).


\bibitem{MV19} B. Musto and  J. Vicary,
                  Orthogonality for Quantum Latin Isometry Squares,
                  EPTCS {\bf 287}, 253 (2019).

\bibitem{Ri20} A. Rico,
Absolutely maximally entangled
states in small system sizes,
\href{https://diglib.uibk.ac.at/ulbtirolhs/content/titleinfo/5327562}{Master Thesis,}
 Innsbruck (2020).

\bibitem{Table_AME} F.~Huber and N.~Wyderka, 
\href{http://www.tp.nt.uni-siegen.de/+fhuber/ame.html}{Table of {A}{M}{E} states},
2021.

\bibitem{yu2020complete}
X.-D. Yu, T.~Simnacher, N.~Wyderka, H.~C. Nguyen, and O.~G{\"u}hne,
A complete  hierarchy for the pure state marginal problem in quantum mechanics,
   {\em Nature Communications} {\bf 12}, 1012  (2021).


\bibitem{BZ17} I. Bengtsson and K. {\.Z}yczkowski, 
{\sl  Geometry of Quantum States}. 2 Ed., Cambridge (2017).

\bibitem{MultiUnitary}
D.~Goyeneche, D.~Alsina, A.~Riera, J.~Latorre, and K.~{\.Z}yczkowski,
Absolutely  maximally entangled states, combinatorial designs and multi-unitary
  matrices,
   {\em Phys. Rev.} { A 92}, 032316 (2015).


\bibitem{SAA2020} S.~A. Rather, S.~Aravinda, and A.~Lakshminarayan,
Creating ensembles of dual
  unitary and maximally entangling quantum evolutions,
   {\em Phys. Rev. Lett.} {\bf 125}, 070501 (2020).
   
\bibitem{BNechita}
T. Benoist and I. Nechita,
On bipartite unitary matrices generating subalgebra-preserving quantum operations,
{\sl Linear Algebra and Appl.} {\bf 521}, 70 (2017).
    
    
\bibitem{Cross_2009} A.~Cross, G.~Smith, J.~A. Smolin, and B.~Zeng, 
Codeword stabilized quantum  codes,
 {\em IEEE Transactions on Information Theory}, 
  {\bf 55}, 433  (2009).

\bibitem{BurchardtRaissi20}
A.~Burchardt and Z.~Raissi, Stochastic local operations with classical
  communication of absolutely maximally entangled states,
   {\em Phys. Rev. A}, {\bf 102}, 022413 (2020).

\bibitem{NonadditiveCodes}
S.~Yu, Q.~Chen, C.~H. Lai, and C.~H. Oh, Nonadditive quantum error-correcting
  code, {\em Phys. Rev. Lett.} {\bf 101}, 090501 (2008).

\bibitem{Rains1999NonbinaryQC}
E.~M. Rains, Nonbinary quantum codes, {\em IEEE Trans. Inf. Theory},
  {\bf 45}, 1827 (1999).

\bibitem{Huber_2020}
F.~Huber and M.~Grassl, Quantum codes of maximal distance and highly
  entangled subspaces, {\em Quantum}, {\bf 4}, 284 (2020).

%
%
%
%
%
%
%

\end{thebibliography}

\begin{thebibliography}{99}

\bibitem{Scott04}  A.~J. Scott, 
Multipartite entanglement, quantum-error-correcting codes, and
  entangling power of quantum evolutions,
   {\em Phys. Rev.} {\bf  A 69}, 052330 (2004).


\bibitem{HelwigAME}
W.~Helwig, W.~Cui, A.~Riera, J.~Latorre, and H.-K. Lo, 
 Absolute maximal  entanglement and quantum secret sharing,
   {\em Phys. Rev.} {\bf  A 86},  052335  (2012).
   
\bibitem{Table_AME} F.~Huber and N.~Wyderka, 
\href{http://www.tp.nt.uni-siegen.de/+fhuber/ame.html}{Table of {A}{M}{E} states},
\newblock Accessed: July 2021.

\bibitem{MazurekGrudka}
P.~Mazurek, M.~Farkas, A.~Grudka, M.~Horodecki, and
  M.~Studzi\ifmmode~\acute{n}\else \'{n}\fi{}ski,
  Quantum error-correction codes and absolutely maximally entangled states,
   {\em Phys. Rev.} {\bf A 101}, 042305  (2020).
   
\bibitem{MultiUnitary}
D.~Goyeneche, D.~Alsina, A.~Riera, J.~Latorre, and K.~{\.Z}yczkowski,
Absolutely  maximally entangled states, combinatorial designs and multi-unitary
  matrices,
   {\em Phys. Rev.} { A 92}, 032316 (2015).

\bibitem{BKP2019} B. Bertini, P. Kos, and T. Prosen,
Operator Entanglement in Local Quantum 
                Circuits I: Chaotic Dual-Unitary Circuits,
                SciPost Phys. 8, 067 (2020).
                
\bibitem{SAA2020} S.~A. Rather, S.~Aravinda, and A.~Lakshminarayan,
Creating ensembles of dual
  unitary and maximally entangling quantum evolutions,
   {\em Phys. Rev. Lett.} {\bf 125}, 070501 (2020).


\bibitem{Pastawski2015HolographicQE}
F.~Pastawski, B.~Yoshida, D.~C. Harlow, and J.~Preskill, 
Holographic quantum
  error-correcting codes: toy models for the bulk/boundary correspondence,
  {\em Journal High Energy Phys.}, {\bf 2015}, 1 (2015).

\bibitem{GRMZ18} D.~Goyeneche, Z.~Raissi, S.~Di~Martino, 
and K.~{\.Z}yczkowski, 
Entanglement and quantum combinatorial designs,
 {\em  Phys. Rev.} {\bf  A 97},  062326 (2018).


\bibitem{MV19} B. Musto and  J. Vicary,
                  Orthogonality for Quantum Latin Isometry Squares,
                  EPTCS {\bf 287}, 253 (2019).

\bibitem{Ri20} A. Rico,
Absolutely maximally entangled
states in small system sizes,
\href{https://diglib.uibk.ac.at/ulbtirolhs/content/titleinfo/5327562}{Master Thesis,}
 Innsbruck (2020).

\bibitem{BNechita}
T. Benoist and I. Nechita,
On bipartite unitary matrices generating subalgebra-preserving quantum operations,
{\sl Linear Algebra and Appl.} {\bf 521}, 70 (2017).

\bibitem{CDN20}
 G. De las Cuevas, T. Drescher and T. Netzer,
  Quantum magic squares: dilations and their limitations,
  {\sl J. Math. Phys.} {\bf  61}, 111704 (2020).


 \bibitem{files} Consult mathematical files
 \href{ https://chaos.if.uj.edu.pl/~karol/Maestro7/data2}{{\sl available here}};
  A. Rico, 
 \href{ https://chaos.if.uj.edu.pl/~karol/Maestro7/data2}{{\sl unpublished note,} 2021}

\bibitem{Huber_2018}
F.~Huber, C.~Eltschka, J.~Siewert, and   O.~G{\"u}hne,  Bounds on absolutely
  maximally entangled states from shadow inequalities, and the quantum
  {MacWilliams} identity, {\em J. Phys.} {\bf 51}, 175301 (2018). 

\bibitem{Zanardi_2001} P. Zanardi,
Entanglement of quantum evolutions,
 {\sl  Phys. Rev.} {\bf  A 63}, 040304 (2001).
 
\bibitem{BZ17} I. Bengtsson and K. {\.Z}yczkowski, 
{\sl  Geometry of Quantum States}. 2 Ed., Cambridge (2017).

\bibitem{bhargavi2017}
B.~Jonnadula, P.~Mandayam, K.~\ifmmode~\dot{Z}\else \.{Z}\fi{}yczkowski, and
  A.~Lakshminarayan, Impact of local dynamics on entangling power, {\em
  Phys. Rev.} {\bf A 95},  040302  (2017).
  
 
\bibitem{Clarisse_2005}
L.~Clarisse, S.~Ghosh, S.~Severini, and A.~Sudbery,
Entangling power of  permutations,
 {\em Phys. Rev.}, {\bf A 72} (2005).
 
 
\bibitem{HIGUCHI2000}
A.~Higuchi and A.~Sudbery, 
How entangled can two couples get?, 
{\em Phys.  Lett.} {\bf  A 273},  213  (2000).
 

\end{thebibliography}

\clearpage
\pagebreak
\newpage

\widetext
\begin{center}
\textbf{\large Supplemental Material for \\
``Thirty-six entangled officers of Euler: Quantum solution to a classically impossible problem''}
\end{center}
\setcounter{equation}{0}
\setcounter{figure}{0}
\setcounter{table}{0}
\makeatletter
\renewcommand{\theequation}{S\arabic{equation}}
\renewcommand{\thefigure}{S\arabic{figure}}
\renewcommand{\bibnumfmt}[1]{[S#1]}
\renewcommand{\citenumfont}[1]{S#1}

\onecolumngrid
\renewcommand{\theequation}{S\arabic{equation}}

\section{Orthogonal Quantum Latin Squares and $4$-partite entangled states.}
\label{AppA}

In this section, we collect the  definitions introduced
in the literature in various contexts\renewcommand{\thefigure}{2a}
and demonstrate the equivalence between different notions used in the main 
body of the work.

{\bf Definition 1.} AME state \cite{Scott04,HelwigAME}.
{\sl A pure quantum state,
 $\ket{\psi} \in \mathcal{H}_d^{\otimes N}$ 
of $N$ parties, each of a local dimension $d$,
 is called absolutely maximally entangled (AME),
 written as} $|\text{AME}(N,d)\rangle$,
  {\sl if it is maximally entangled for every bipartition, i.e. the partial trace $Tr_S \ket{\psi}\bra{\psi} \propto \mathbb{I}$, for any subsystem $S$ of $|S|=\lfloor N/2 \rfloor$ parties. }

\smallskip

In this work, we analyze the case of a fourpartite state, $N=4$.
A list of  known AME states is available  at \cite{Table_AME}.
In general, an AME state consisting of an even number $N$ of
subsystems with $d$ levels each
leads to a  pure quantum error correction code $(\!(N,1,N/2+1)\!)_d$, 
 which saturates the Singleton bound \cite{MazurekGrudka}.

\medskip

{\bf Definition 2.} Multiunitary matrix \cite{MultiUnitary}.
{\sl A unitary matrix $U$ of order $d^2$ is called $2$-unitary
if the reshuffled matrix $U^{\R}$ and partially transposed matrix $U^{\PT}$ are also unitary.}

\smallskip

Matrices with the last condition relaxed also play a role
in studies on many-body quantum dynamics: 
A unitary matrix $U$ of size $d^2$, for which $U^{\R}$ is unitary,
is called {\sl dual-unitary} \cite{BKP2019,SAA2020}.
In general,
 a matrix $U$ of order $d^k$ is called $k$-unitary (or multiunitary),
if it remains unitary for all $(2k-1)(k-1)$ reorderings of $2k$ indices
which define the matrix.
Any $2$-unitary matrix $U$ of size $d^2$describes a  bipartite quantum gate, 
which maximizes the entangling power, $e_p(U)=1$, see Supplementary Material Section 3.

\medskip

{\bf Definition 3.} Perfect tensor \cite{Pastawski2015HolographicQE}.
{\sl A tensor $T_{i_1\dots i_{2k}}$
with $2k$ indices, each running from $1$ do $d$
is called {\sl perfect},
if any of its flattening into a matrix of order 
$d^k$ is unitary.}

\medskip

The notion of orthogonal quantum Latin squares (OQLS)
was introduced in \cite{GRMZ18,MV19}.
Here we follow an alternative  approach of Rico \cite{Ri20}
to assure full consistency with $2$-unitarity.
Let $|\chi^+\rangle=\sum_{i=1}^d |i,i\rangle$
denote the (non-normalized) maximally entangled Bell state
of a   $2$-qudit system.
 Any bipartite quantum state,
$|\psi\rangle = \sum_{k,\ell=1}^d C_{k,\ell} |k,\ell\rangle$,
can be also written as
 $|\psi\rangle =  (C \otimes {\mathbbm I}) |\chi^+\rangle$.

\smallskip

{\bf Definition 4.}  OQLS.
{\sl Consider a  set of $d^2$ bipartite states in 
$\mathcal{H}_d \otimes \mathcal{H}_d$,
which can be written in a product basis,
$|\psi_{ij}\rangle = \sum_{k,\ell=1}^d C^{i,j}_{k\ell} |k,\ell\rangle$
for $i,j=1,\dots,d$.
Such a set forms an OQLS if
(a) the states satisfy ortogonality relations,
$\langle\psi_{ij}|\psi_{k \ell}\rangle = 
 \delta_{ik}\delta_{j\ell}$;
and the block matrix $\tilde C$ of size $d^2$
written
$\tilde C=(C^{1,1},\dots ,  C^{1,d};
\dots ;
  C^{d,1},\dots  , C^{d,d})$  is block unitary, 
  so that the conditions
  (b) 
$\sum_{i=1}^d C^{i,j} (C^{i,\ell})^{\dagger} = \delta_{j,\ell}{\mathbbm I}$,
and
(c)  
$\sum_{j=1}^d C^{i,j} (C^{k,j})^{\dagger} =  \delta_{i,k}{\mathbbm I}$,
are satisfied.
}

\smallskip
Observe that the orthogonality relations, equivalent to
${\rm Tr} \; C^{i,j} (C^{k,\ell})^{\dagger}=
 \delta_{ik}\delta_{j\ell}$,
 correspond to the
 `different location' conditions
  for strong sudoku satisfied by the 2-unitary permutations
  presented in Eq.~\ref{P9} and \ref{P9R}.
  Furthermore, note that the block unitarity of $\tilde C$
implies that the related block matrix, 
$\tilde B$ 
with blocks $B^{i,j}=C^{i,j} (C^{i,j})^{\dagger}$ 
  is block bistochastic,
  $\sum_{i=1}^d B^{i,j} ={\mathbbm I}$,
and
   $\sum_{j=1}^d  B^{i,j} ={\mathbbm I}$.
   Hence any OQLS described by the block matrix  $\tilde C$
   implies a block bistochastic matrix $\tilde B$ introduced in \cite{BNechita}, and
   recently studied in \cite{CDN20}
   under the name of {\sl quantum magic square}.
 
\medskip

To show these notions in action, we shall analyze the case 
of a Graeco-Latin square (OLS) of size three, shown in Fig. 2.
To get a set of $9$ bipartite vectors $|\psi_{ij}\rangle$
in a given cell   $(A,\alpha)$ 
we replace the classical symbols
by a  bipartite quantum state  $|A,\alpha\rangle$.
Thus each state has a product form,
$|\psi_{i,j}\rangle = |i\oplus j\rangle \otimes |i \oplus 2j \rangle$, 
which is a consequence of the fact that OLS are classical. Notice that the operations inside kets are performed  modulo $3$.

The same information is encoded in the
tensor determined by
\begin{equation}
\label{3perfect}
 T_{ij k \ell}=  
 \delta_{k,i \oplus j}\; \delta_{\ell,i \oplus 2j}, 
\end{equation}
with addition operations performed  modulo $3$.
It is easy to check that this tensor is perfect.
Furthermore, the corresponding state
\begin{equation}
\label{ame43}
|\text{AME}(4,3)\rangle=
\frac{1}{3}\sum_{i,j=1,2,3} |i\rangle|j\rangle |i \oplus j\rangle |i \oplus 2 j\rangle 
\end{equation}
belongs to AME states, as 
all reduced density matrices to two qutrits are equal to ${\mathbbm I}_9/9$. 
One may present the tensor $ T_{ij k \ell}$ in a form of $9\times 9$ permutation matrix $P_9$, where non-vanishing element of $ T_{ij k \ell}$ corresponds to a non-zero entry of $P_9$ on the intersection of $j+3(i-1)  $-th row with $\ell +3(k-1)$-th column. 
In other words,  the corresponding permutation matrix
\begin{equation}
\label{P9}
P_9=
\left(\begin{array}{ccc|ccc|ccc}
0&0&0&0&0&1&0&0&0\\
0&0&0&0&0&0&0&1&0\\
1&0&0&0&0&0&0&0&0\\
\hline
0&0&0&0&0&0&1&0&0\\
0&0&1&0&0&0&0&0&0\\
0&0&0&0&1&0&0&0&0\\
\hline
0&1&0&0&0&0&0&0&0\\
0&0&0&1&0&0&0&0&0\\
0&0&0&0&0&0&0&0&1\\
\end{array}\right) 
\end{equation}
is 2-unitary, since 
 the reshuffled and partially transposed matrices
\begin{equation}
\label{P9R}
P_9^{\R}=
\left(\begin{array}{ccc|ccc|ccc}
0&0&0&0&0&0&1&0&0\\
0&0&1&0&0&0&0&0&0\\
0&0&0&0&1&0&0&0&0\\
\hline
0&0&0&0&0&1&0&0&0\\
0&0&0&0&0&0&0&1&0\\
1&0&0&0&0&0&0&0&0\\
\hline
0&1&0&0&0&0&0&0&0\\
0&0&0&1&0&0&0&0&0\\
0&0&0&0&0&0&0&0&1\\
\end{array}\right)  \qquad \text{and}
\qquad
P_9^{\PT}=
\left(\begin{array}{ccc|ccc|ccc}
0&0&1&0&0&0&0&0&0\\
0&0&0&0&0&0&0&1&0\\
0&0&0&1&0&0&0&0&0\\
\hline
0&0&0&0&0&0&1&0&0\\
0&0&0&0&0&1&0&0&0\\
0&1&0&0&0&0&0&0&0\\
\hline
0&0&0&0&1&0&0&0&0\\
1&0&0&0&0&0&0&0&0\\
0&0&0&0&0&0&0&0&1\\
\end{array}\right) 
\end{equation}
also form permutation matrices. 
Observe that positions of non-zero entries obey the rules
of a strong Sudoku: in each row, column and block there is a single
entry equal to $1$. Furthermore, all the
locations of these entries  in each block are different.
\medskip

Let us return now to the general case of an arbitrary dimension $d$
and relax the assumption that the $2$-unitary matrix of order $d^2$
is a permutation. To demonstrate relations 
between different notions introduced independently in different communities 
we recall the equivalence relations.

\medskip
{\bf Proposition 1}. 
{\sl The following statements are equivalent:}
\begin{enumerate}
{\sl
\item There exist orthogonal quantum Latin squares of size $d$, 
\item There exists an AME state of four subsystems with $d$ levels each,
\item There exists a 2-unitary matrix $U$ of size $d^2$, 
\item There exists a perfect tensor with four indices, each running from $1$ to $d$.}
\end{enumerate}
\medskip 
To show that this is the case, note first that definition~3 of a perfect tensor 
is equivalent to definition~2 of a multiunitary matrix.
Furthermore, these properties imply that the state given by Eq. 1
has all reductions of size two maximally mixed,
 so  according to definition~1 it forms an AME state 
 $|\Psi_{ABCD}\rangle$ of four parties.
The three constraints, (a), (b) and (c) in definition~4 
are equivalent to the fact that partial traces over
selected subsystems, $CD$, $BD$ and $AD$, respectively
are maximally mixed, 
so existence of  OQLS and an AME state of the corresponding system, 
is equivalent. 

Observe that a matrix $U$ is 2-unitary if and only if the corresponding matrix $U^{\R}$, or equivalently $U^{\PT}$, is 2-unitary. 
Indeed, all three corresponding states are related by the permutation of subsystems. 
Note that conditions imposed on AME states are invariant under the permutation of subsystems. 


Furthermore, a 2-unitary matrix $U$ remains 2-unitary after any local operation, i.e.  
$U \mapsto \tilde{U} (U_A\otimes U_B)U(U_C\otimes U_D)$ where 
$U_A,U_B,U_C,U_D \in \mathbb{U}(6)$ are unitary matrices of size $6$. 
Thus, by this mean one can entangle officers from a classical OLS, creating \emph{apparently} quantum solution of OQLS.
By apparent, we express the fact that it can be disentangled to a classical OLS by applying reverse local operation $(U_A^\dagger\otimes U_B^\dagger)\tilde{U}(U_C^\dagger\otimes U_D^\dagger)$.
Observe that a similar disentangling procedure is not feasible in the case of any AME$(4,6)$ state, since it could not have been produced using classical designs. Consequently, we call our result a \emph{genuinely} quantum solution to 
OQLS of size $d=6$.

Note that the state AME$(4,3)$
corresponds to a classical OLS, as the $2$-unitary matrix
is a permutation $P_9$, so the entries of the design are product states.
We are not aware whether for $d=3$
there exist genuinely quantum orthogonal Latin squares.
However, such a design exists for $d=6$,
as will be demonstrated below.

\section{Proof of Theorem 1 on the existence of AME$(4,6)$ }
\label{AMEcheckApp}

In this section we prove Theorem 1 by construction. We introduce a matrix $\mathcal{U}$ of size $36$ and show that it is $2$-unitary.
Its non vanishing entries, determined by the tensor, $\mathcal{U}_{p,s}=T_{ijkl}$, with $p=j+6(i-1)$ and $s=\ell(k-1)$, are shown in Fig 3.
To show $2$-unitarity of the matrix $\mathcal{U}$ (also denoted $\mathcal{U}_{36}$ in the main text)
 we verify that three related matrices: $\mathcal{U}$, $\mathcal{U}^{\R}$ and $(\mathcal{U}^{\R})^{\PT}$ are unitary. 
By definition~2, this implies that the matrix $\mathcal{U}^{\R}$ is $2$-unitary, which is equivalent to the fact that the matrix $\mathcal{U}$ is $2$-unitary. 
Each of those matrices has the structure of nine $4\times 4$ blocks. 
Hence our task simplifies to verification that constituent blocks are unitary matrices. 

\begin{figure}[h!]
\renewcommand{\thefigure}{S1}
\centering
\includegraphics[scale=0.35]{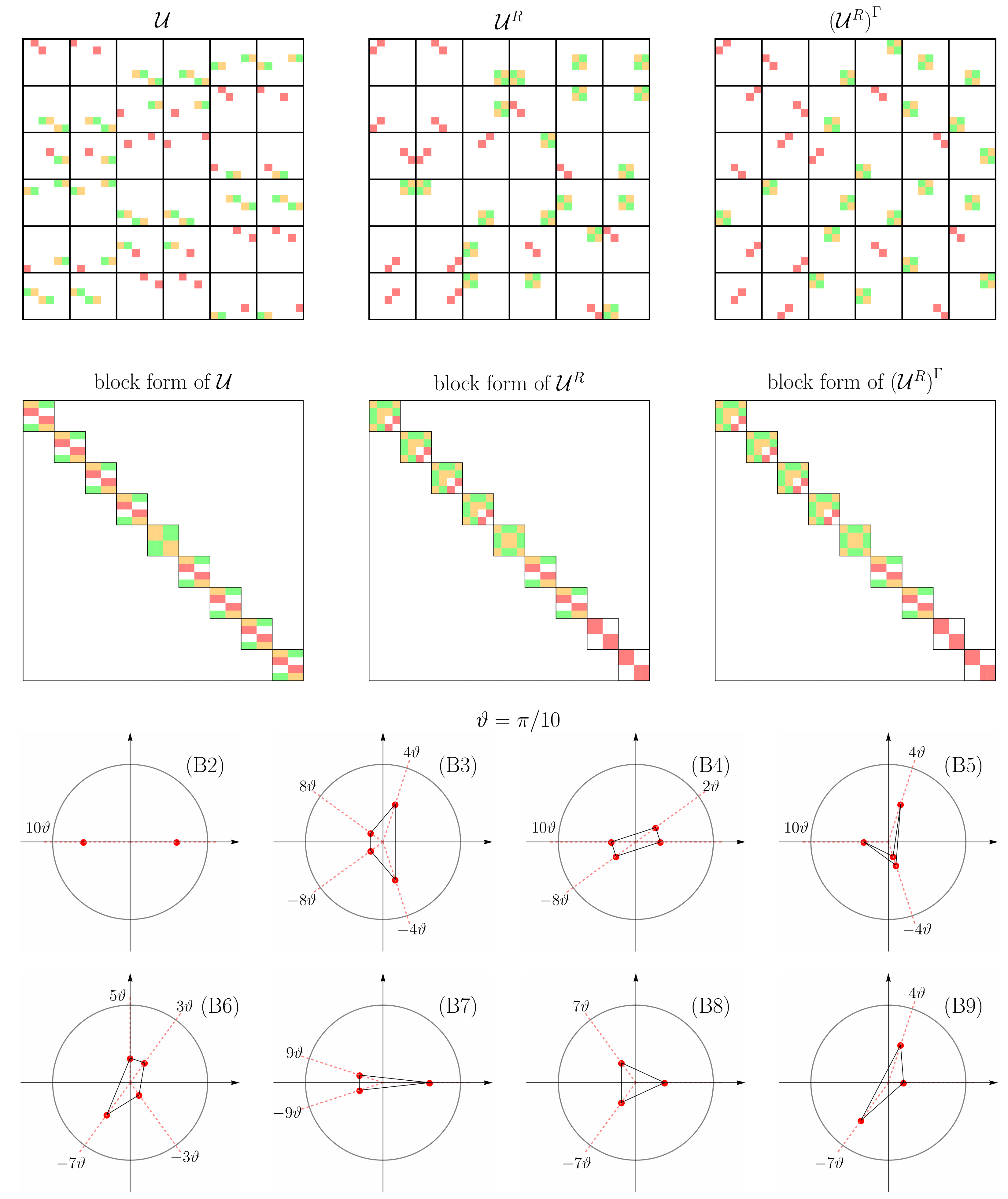}
\caption{Structure of three related matrices $\mathcal{U}$, $\mathcal{U}^{\R}$ and  $(\mathcal{U}^{\R})^{\PT}$ of order 36 are presented on top. 
The modulus of a non-vanishing element is represented by the intensity of the background color. 
Note that similarly to rows and columns, blocks also satisfy orthogonality conditions b) and c) from definition $4$ of OQLS.
Each matrix has the structure of nine $4\times 4$ blocks, the structure within the blocks, however, is different for matrix $\mathcal{U}$ and matrices $\mathcal{U}^{\R}$, $(\mathcal{U}^{\R})^{\PT}$. 
Orthogonality between pairs of rows in block matrices might be presented as a constellation of two, three, or four points on the complex plane which sum up to zero. 
Constellations related to Eqs.~\ref{constell}-\ref{constell2} are indicated. 
In most cases, constellations are simply pairs or double pairs of antipodal points, which clearly sum up to zero. 
There are six non-trivial constellations, all of them are depicted. Note that all phases are multiples of $\vartheta = \pi/10$.}
\label{fig6}
\end{figure}

\begin{figure}[h!]
\renewcommand{\thefigure}{S2}
\includegraphics[scale=0.85]{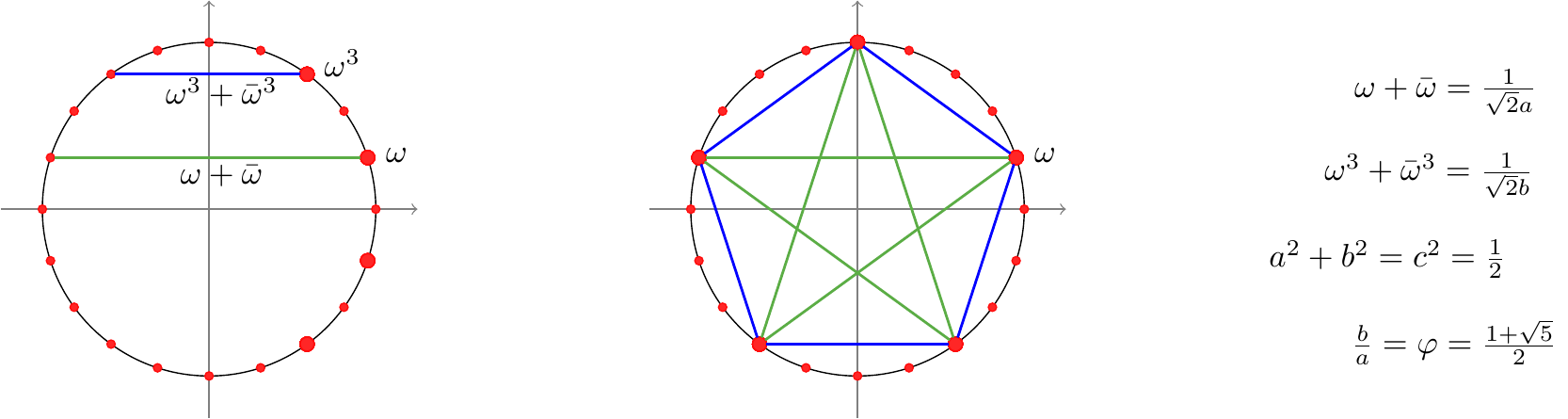}
\caption{Geometric interpretation of the constants $a,b$ and $c$
 appearing in  Eq. 4 with use of a regular pentagon. First two numbers contain the phase $\omega=\exp(i\pi/10)$
and are related to the golden ratio $\varphi$. }
\label{penta}
\end{figure}

Consider the $2$-unitary matrix $\mathcal{U}$. 
Interestingly, except for one block component in the matrix $\mathcal{U}$,
all eight block components are equivalent (up to a multiplication of rows and columns by adequate phases) to the following $4\times 4$ unitary matrix:
\[V=
\begin{bmatrix}
a & a & b & b\\
0 & 0 & c & -c\\
c& -c & 0 & 0\\
b & b & -a & -a
\end{bmatrix}.
\]
Orthogonality relations between rows in the matrix above might be presented as pairs of antipodal points on the complex plane, for example, orthogonality between the first two rows reads
\begin{equation}
\label{constell}
bc \big(1-1\big)=0.
\end{equation} 
The exceptional block of matrix $\mathcal{U}$ is presented on the right top corner in Fig. 3.  
Six orthogonality relations between rows read
\begin{align}
a^2\big(\omega^{8}  +\omega^{-8}\big) + b^2 \big(\omega^{4} +\omega^{-4}\big)&=0, \\
a b \big( 1  +\omega^{2} +  \omega^{-8} -1\big)&=0,  \\ \nonumber
a b \big( \omega^{-2}  +\omega^{2} +  \omega^{-8} + \omega^{8}\big)&=0,
\end{align}
up to a phase factor,
with $\omega=\exp(i \pi/10)$. 
Each equation might be presented as a unitarity rectangle - a constellation of four points in the complex plane which sum up to zero, as it is shown on Fig. ~\ref{fig6}. 
Observe that the second and third equations above are relevant to two pairs of antipodal points on the complex plane. 
Geometric interpretation of the numbers $a$ and $b$
is shown in Fig.~\ref{penta}.



Although the matrices $\mathcal{U}^{\R}$ and $ (\mathcal{U}^{\R})^{\PT}$ enjoy the structure of nine $4\times 4$ blocks,  similar to $\mathcal{U}$,
the particular arrangement inside their blocks is significantly different from the $\mathcal{U}$ matrix. 
Blocks in $\mathcal{U}^{\R}$ and $(\mathcal{U}^{\R})^{\PT}$ matrices are of four distinct types up to multiplication of their rows and columns by phase factors, see Fig.~\ref{fig6}. 
Orthogonality relations between rows of both matrices reflect their diversed structure. 
In particular, we distinguished five additional orthogonality relations given by the following equations:
\begin{align}
a^2 \omega^{4} +ab \big(\omega^{10}  +\omega^{-4}\big) + b^2 \omega^{-4} &=0, \\
a^2 \omega^{-3} +ab \big(\omega^{5}  +\omega^{3}\big) + b^2 \omega^{-7} &=0, \\
ab \big(\omega^{-4}  +\omega^{-6}\big) +  bc \omega^{5}  &=0, \\
ab \big(\omega^{-8}  +\omega^{-2}\big) +  ac \omega^{5}  &=0, \\
\label{constell2}
a^2 + b^2 \omega^{4} + bc\omega^{-7} &=0.
\end{align}
Related constellations are presented on Fig.~\ref{fig6}. 
The above-listed equalities provide orthogonality between rows in the three matrices $\mathcal{U}$, $\mathcal{U}^{\R}$ and $(\mathcal{U}^{\R})^{\PT}$.

Observe that in the presented solution
 each officer is entangled with no more than $3$ officers
 out of remaining $35$. This  implies the $2$-unitary matrix $\mathcal{U}$ is sparse.  
 An explicit form of the matrix $\mathcal{U}$ is available on line in several formats
 \cite{files}, together with an explicit form of the corresponding AME$(4,6)$
 state determined by $\mathcal{U}$ and Eq. 3.
 Further numerical results  suggest that any $2$-unitary matrix 
  $\mathcal{U}_{36}$ has complex entries. Therefore, it is tempting to conjecture  that
  there exists no solution in the set $\mathbb{O}(36)$ of orthogonal matrices of this size. 
  
Three amplitudes $a,b,c$ which appear in the presented construction might be defined as the unique solution of the following algebraic equations: $a^2+b^2 =c^2 =1/2$ and $b/a = \varphi = (1+\sqrt{5} )/2$, see Fig.~\ref{penta}. 
Notice the similarities between algebraic equations which lead to values $a,b,c$, and the algebraic equations which lead to the amplitudes in a heterogenous AME state in $2\times 3\times 3\times 3 $ system presented in \cite{Huber_2018}. 

The phases of the coefficients shown in Fig. 3 of the main body of the
paper, being multiples of $\omega=\exp(i \pi/10)$, 
 are chosen in such a way that 
  all  $36$ quantum states $\ket{\psi_{ij}}$, each represented by a single
   row of the 2-unitary matrix $\mathcal{U}$,
  are equivalent to the standard, two-qubit Bell state.
 This is fact easy to see for any state 
 formed by two coefficients of moduli, $|c|=1/\sqrt{2}$,
 as states $\ket{\psi_{11}}$ or $\ket{\psi_{56}}$ represented in the second and the third line in
the upper left block in Fig. 3 respectively.
To show that this property holds also for other states, 
note that the state $\ket{\psi_{63}}$ corresponding to the first line of
the aforementioned block can be written in the product basis as
\[
\ket{\psi_{63}} =
a\omega^{10}\ket{{11}}+
b\omega^{15}\ket{{12}}+
b\omega^{5}\ket{{21}}+
a\ket{{22}}.
\]
Thus the partial trace of the projector reads, 
${\rm Tr}_{\rm B} \ket{\psi_{63}}\bra{\psi_{63}}=
{\rm diag}(a^2+b^2, a^2+b^2)={\mathbb I}/2$.
This proves that  $\ket{\psi_{63}}$ is 
locally equivalent to the maximally entangled Bell state.
A similar reasoning works for all other states consisting of four terms
and represented in Fig. 3 by green and yellow elements. 
Hence all $36$ states, 
corresponding to $36$ entangled officers of Euler,
can be considered as  maximally entangled, two qubit states.

\section{Generating 2-unitary matrix of order 36 using a dynamical map}
\label{SM}

{\sl Entangling power}  of a bipartite unitary gate $U$ 
is defined as the mean entanglement produced
by the gate,
$e_p(U)=C_d \,\overline{ \mathcal{E}(|\psi_{AB}\rangle)}$
with
$|\psi_{AB}\rangle = U\left(|\phi_A\rangle \otimes |\phi_B\rangle\right)$.
The average, indicated by the overline, is taken over the Haar measure of random states in each subsystem \cite{Zanardi_2001} and we choose the normalization $C_d=(d+1)/(d-1)$ so that the maximum value of the entangling power is $1$.
As a measure of entanglement it is convenient to choose
the linear entropy of the reduced density matrix, 
$\mathcal{E}(\ket{\psi}_{AB})=1-\text{Tr}(\rho_{A}^2)$,
where $\rho_A={\rm Tr}_B|\psi_{AB} \rangle \langle \psi_{AB}|$.
An alternative approach to measure entanglement in a bipartite gate $U$ 
of order $d^2$ is to
use its operator Schmidt decomposition \cite{BZ17} 
$U= \sum_{j=1}^{d^2} \sqrt{\lambda_j} \, {A_j} \otimes {B_j}$,
where the matrices ${A_j}$ and ${B_j}$
form an orthonormal basis in the space of operators,
while the Schmidt coefficients $\lambda_j$
are given by squared singular values of the reshuffled matrix $U^{\R}$
-- see Eq. 2. 
To quantify non-locality of the gate one uses
the {\sl operator entanglement}, 
defined by the linear entropy of the Schmidt vector,
$E(U)=1- \big(\sum_{j=1}^{d^2} \lambda_j^2\bigr)/d^4$.
It is convenient to introduce 
the SWAP operator $S$, defined by the relation
$S\left(|\phi_A\rangle \otimes|\phi_B\rangle\right) =|\phi_B\rangle \otimes|\phi_A\rangle$.
Note that $S^{\R}$ is unitary, which implies that $E(S)=1-1/d^2$.
It allows one to show a direct link between both quantities and
express the entangling power of $U$
in terms of the operator entanglement  \cite{Zanardi_2001} ,
\begin{equation}
e_p(U)=\frac{1}{E(S)}\Big(E(U)+E(US)-E(S)\Big),
\label{eq:epdef}
\end{equation}
which implies the normalization, $0 \leq e_p(U) \leq 1$.
 Lower bound, $e_p(U)=0$  is saturated by any local gate of a product form, 
 $U=u_A \otimes u_B$,
 and the SWAP gate $S$.

\begin{figure}[h!]
\renewcommand{\thefigure}{S3}
\centering
	\begin{tikzpicture}
		\node (a) at (-3,0) {$\tilde{P}$ =
\setlength{\tabcolsep}{3pt}
\renewcommand{\arraystretch}{1.4}
\begin{tabular}{|llllll|}
\hline
$11$ & $22$ & {\color{red} $33$}  & {\color{red} $44$} & $55$ & $66$ \\ 
$23$ & $14$ & $45$ & $36$ & $61$ & $52$ \\ 
$32$ & $41$ & $64$ & $53$ & $16$ & $25$ \\ 
$46$ & $35$ & $51$ & $62$ & $24$ & $13$ \\ 
$54$ & $63$ & $26$ & $15$ & $42$ & $31$ \\ 
$65$ & $56$ & $12$ & $21$ & {\color{red} $33$} & {\color{red} $44$} \\
\hline
\end{tabular}
 };
    \node (a) at (-0.5,0) {=};
    \node (a) at (3.3,0) {\includegraphics[scale=0.95]{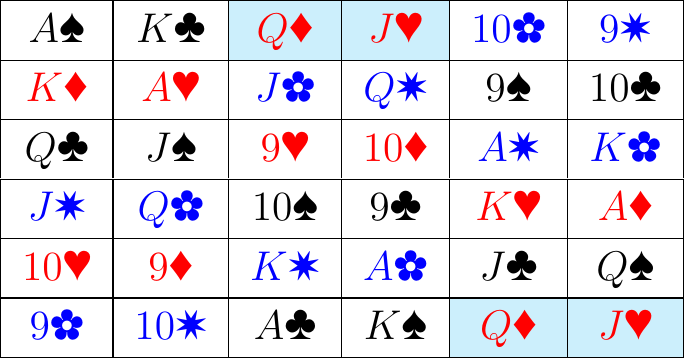}};
	\end{tikzpicture}
	\caption{The design in $d=6$ that is the closest possible to being an OLS \cite{Clarisse_2005}. The two pairs of marked entries are repeated and not all possible 36 pairs are found. }\label{fig36c}
\label{P_tilde}
\end{figure}  
  
  Entangling power does not distinguish between locally inequivalent 
  gates like $U$ and $US$, as $e_p(U)=e_p(US)$. To distinguish them 
 one can use a complementary quantity called 
 {\sl gate typicality} \cite{bhargavi2017},
\begin{equation}
g_t(U)=\frac{1}{2E(S)}\Big(E(U)-E(US)+E(S)\Big),
\label{eq:epdef}
\end{equation}
satisfying  $0 \leq g_t(U) \leq 1$. Lower bound, $g_t(U)=0$, is attained 
 by gates of a product form, while
 the upper bound $g_t(U)=1$ is achieved by the SWAP gate
 and  locally equivalent gates.
  Any $2$-unitary matrix $\mathcal{U}$ is maximally non-local, also if composed with SWAP,
  and one has $E(\mathcal{U})=E(\mathcal{U}S)=E(S)$,
  so that  $g_t(\mathcal{U})=1/2$ and  $e_p(\mathcal{U}) =1$.
Note that there is no 2-unitary matrix of size $d^2=4$,
for which the maximal value $e_p=1$  
of the entangling power is achieved \cite{Zanardi_2001, Clarisse_2005}.
This is equivalent to the fact that 
there are no AME states for a four-qubit system \cite{HIGUCHI2000}.

To look for $2$-unitary matrices of size $d^2$
 using the dynamical map,  $U_{n+1}=\mathcal{M}_{\PT \R}[U_n]$,
  presented in the main body of the text
 one needs to find an appropriate initial matrix.
While for $d=3$ a random unitary matrix of order nine
with a significant probability generates 
a 2-unitary matrix, finding a suitable seed  for $d=6$
is considerably more demanding. It is natural to
consider the design which gives the best approximation to a Graeco-Latin square \cite{Clarisse_2005} and this is shown in Fig.~\ref{P_tilde}.

\begin{figure}[h!]
\renewcommand{\thefigure}{S4}
\centering
	\begin{tikzpicture}
		\node (a) at (-3,0) {$\tilde{P_s}$ =
\setlength{\tabcolsep}{3pt}
\renewcommand{\arraystretch}{1.4}
\begin{tabular}{|llllll|}
\hline
$11$ & $22$ & $33$ & $44$ & ${\color{red} 55}$ & $66$ \\ 
$23$ & $14$ & $45$ & $36$ & $61$ & $52$ \\ 
$32$ & $41$ & ${\color{red} 64}$ & $53$ & $16$ & $25$ \\ 
$46$ & $35$ & $51$ & $62$ & $24$ & $13$ \\ 
${\color{red} 64}$ & $56$ & $26$ & $15$ & {${\blu 4}3$} &  {${\blu 3}1$} \\ 
${\color{red} 55}$ & $63$ & $12$ & $21$ &  {${\blu 4}2$} &  {${\blu 3}4$} \\
\hline 
\end{tabular}
 };
    \node (a) at (-0.5,0) {=};
    \node (a) at (3.3,0) {\includegraphics[scale=0.95]{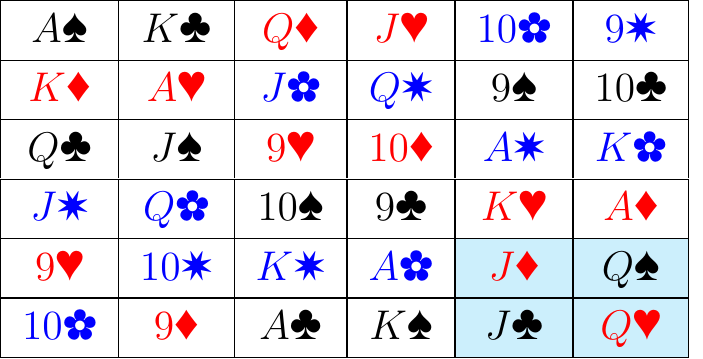}};
	\end{tikzpicture}
	\caption{The design in $\tilde{P}_s$ does not form an
   OLS of dimension 6 since two pair of ranks are in the same columns.
	Its small perturbation used as a seed to the algorithm gives the
	 2-unitary matrix $\mathcal{U}_{36}$ and the desired  state
	    $|\text{AME}(4,6)\rangle$.}
\label{P36b}
\end{figure}

A permutation matrix of order $36$, denoted as $P_{36}$, can be obtained from $\tilde{P}$: if $P_{36}$ is partitioned into $6\times 6$ blocks, the only nonzero $(=1)$ entry of the $ij-{\rm th}$ block is given by the corresponding entry in $\tilde{P}$. For example, the $(5,1)$ block is such that its $4$th column and $5$th row is $1$. This has the maximum entangling power over all permutations of order $36$ and is given by  $e_p(P_{36})=314/315 \approx 0.996825$. If one starts with $P_{36}$ as the seed, then $e_p(U_{n}) \rightarrow e_p(A) = 419/420 \approx 0.9976$ as $n \rightarrow \infty$ where 
$U_{n}=\mathcal{M}_{\PT \R}^n\left[P_{36}\right]$. Interestingly, $A$ is an orthogonal matrix such that $A^{\PT}$ is unitary but $A^R$ fails to be one. 

\begin{figure}[h!]
\renewcommand{\thefigure}{S5}
\centering
		\includegraphics[scale=0.55]{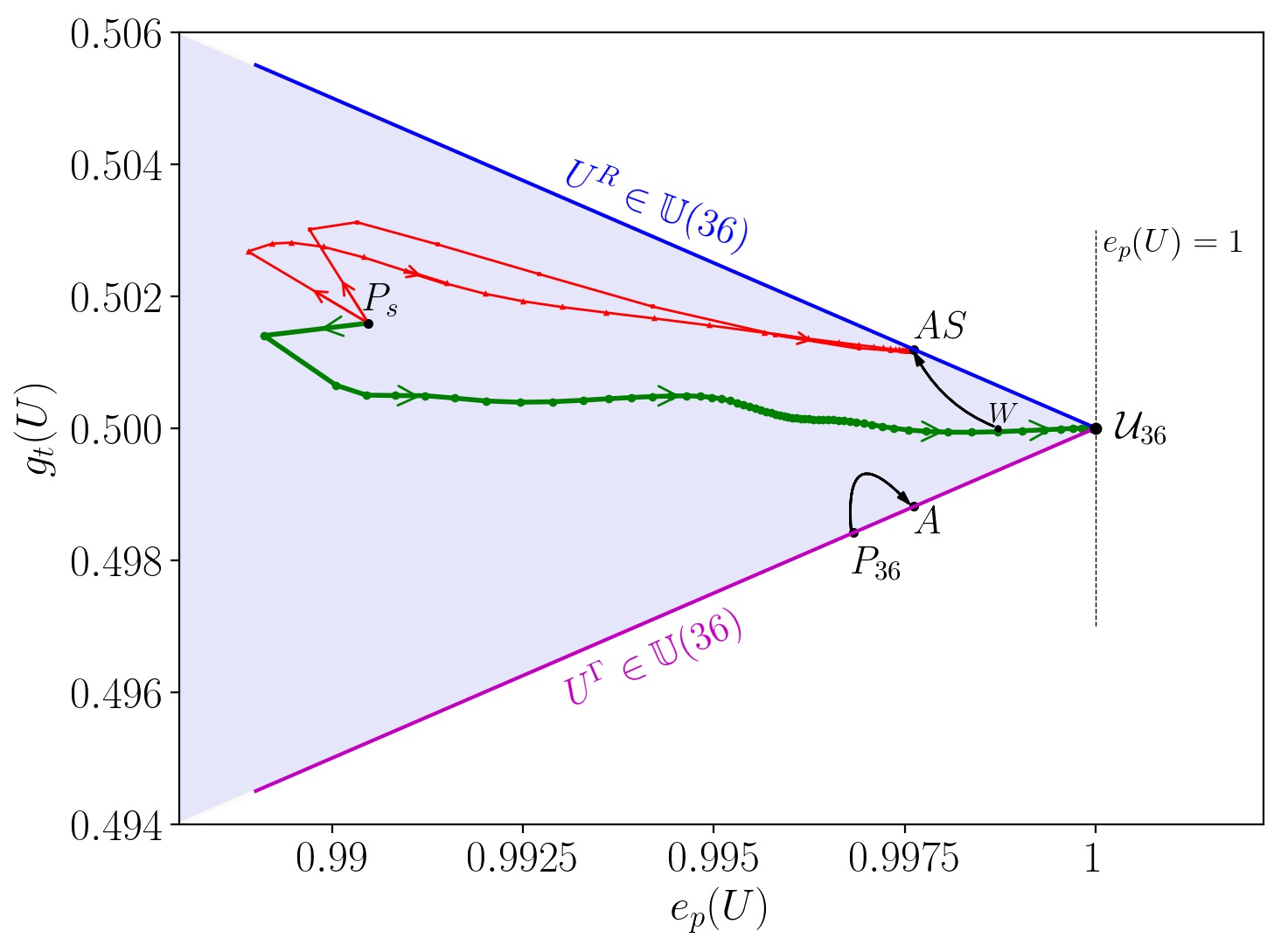}
		\caption{ A sample of trajectories initiated in the neighborhood of $P_s$
and generated by the map    $\mathcal{M}_{\PT \R}$ are shown,
where every third iteration is marked. Many of the trajectories end up at a strongly attracting fixed point labelled $A$ or its partner that is multiplied by the SWAP gate $S$, $A\,S$ while a small fraction are able to reach the 2-unitary fixed point, $\mathcal{U}_{36}$. If the map is initiated from the best approximation to an OLS, $P_{36}$, or from its neighborhood the trajectory ends up mostly at the point corresponding to the matrix $A$,
		which has a larger entangling power than $P_{36}$. If one starts from $W$ which is an orthogonal matrix with an even larger entangling power than $A$, it ends up at the point corresponding to $A\, S$.}
\label{fig:epgtconv}
\end{figure}

 This is an encouraging result and gives a way to explore unitary operators which have entangling powers larger than that of $P_{36}$. Similarly, if we start in the neighborhood of $P_{36}$, {\it i.e.} we take seeds of the form $P_{36}\exp(i\epsilon H)$ where $\epsilon \ll 1$ and $H$ is Hermitian; $H=(M+M^T)/2$ with entries of $M$ sampled from the normal distribution, then the map converges to unitaries with larger values of entangling power. However, $e_p(U_{n}) \approx 0.9991$ is the largest value obtained by us while starting in the vicinity of $P_{36}$ (and for sufficiently large $n$).

In order to generate a $2$-unitary matrix of size $36$, one can search in the vicinity of other permutation matrices.
Since the search space is very large ($36! \approx 10^{41}$), 
we restrict our attention to permutations that are in the vicinity of $P_{36}$. 
One of the permutations that works, denoted below as $P_s$, can be constructed from the design in Fig.~\ref{P36b} as described above.

\begin{figure}[h!]
\renewcommand{\thefigure}{S6}
\centering
		\includegraphics[scale=0.50]{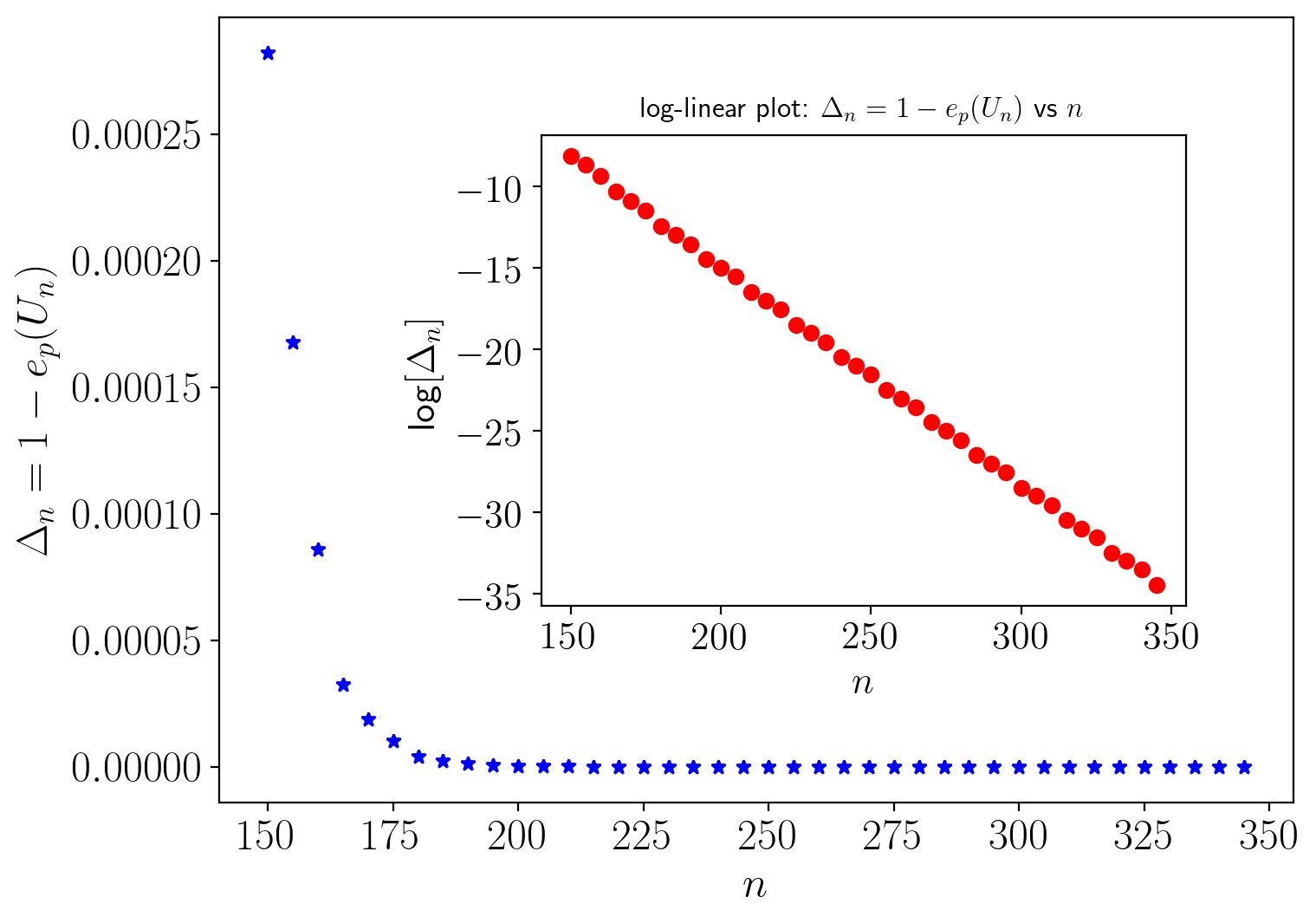}
		\caption{Convergence of a trajectory initiated  in the neighborhood of $P_s$ to a 2-unitary matrix
		 quantified by the deviation $\Delta_n=1-e_p(U_{n})$ 
		 is plotted as a function of the number $n$ of iterations.
		For $n>150$ the deviation  $\Delta_n$ decays exponentially as shown in the inset.}
\label{epconv}
\end{figure}
The matrix  $\tilde{P_s}$ differs from $\tilde{P}$ in the last two rows, see Fig.~\ref{P36b}. The entangling power of the permutation matrix $P_s$ is $e_p(P_s)=104/105 \approx 0.9905$. Much better results can be obtained 
if one starts with a matrix from the neighborhood of $P_s$, as indicated above.
   The dynamical map with a finite probability ($p \approx 6\%$)
    converges to a 2-unitary matrix
    such that $e_p(U_{n})=1$ 
    up to a machine precision for $n \sim 10^3$.
    Trajectories of some initial seed unitaries in the neighborhood of $P_s$ are shown in Fig.~\ref{fig:epgtconv}. Every third iteration is marked. Most of these seed unitaries converge to strongly attracting fixed points of the map, such as  $A$ or $A\, S$ (local extrema) while a few converge to 2-unitaries (global extrema). Nearby initial conditions with almost the same values of entangling power and gate typicality converge to different fixed points and explain the complex dynamics induced by the map on the $(e_p,g_t)$ plane.
     The iteration procedure and convergence
     of the trajectory $U_n$
      to 2-unitary matrix on the $(e_p, g_t)$ plane
      is visualised  in Fig.~\ref{fig:epgtconv}.
      Convergence  speed and the  accuracy obtained is 
       quantitatively described in Fig.~\ref{epconv}.
and implies that 
 after a sufficiently large number of iterations, matrix $U_n$ becomes 2-unitary. 
The output matrix, $\mathcal{U} = U_{n}$, can be put in a block diagonal form 
consisting of three blocks of size $12$.
%
 There are also other ways of getting appropriate seeds to generate 2-unitary operators, such as enphasing of the permutation $P_s$ (equivalent to multiplication by a diagonal unitary). 


\section{An apparent solution of the  {\bf Euler's} problem for $d=6$}
\label{36cube}

\begin{figure}[h!]
\renewcommand{\thefigure}{S7}
\centering
\includegraphics[width=0.30 \textwidth]{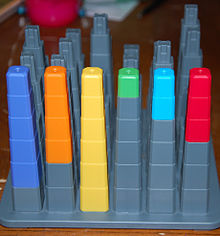}
\hskip 0.4cm
\includegraphics[width=0.288 \textwidth]{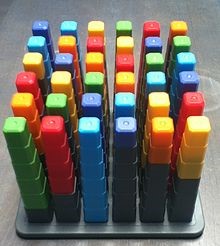}
		\caption{Left panel: Puzzle 36cuBe
		 advertized as
		{\sl the World's Most Challenging Puzzle}
		  consisting of 36 pieces of six colors and six heights
		and (right panel) its apparent solution. 
		Closer inspection of all the pieces
		   reveals that it corresponds to the pattern
		    presented in Fig.~\ref{P36b},
		     as two colums contain two pieces of the same height, in agreement
		       with the conjecture of Euler and the theorem of Tarry.
		     }
\label{figcube36}
\end{figure}

An interesting practical application of the standard Euler's problem 
of $36$ officers is worth to be mentioned here.
A puzzle called {\sl 36cuBe} 
designed in 2008 by D. C. Niederman
is directly linked to this mathematical question:
the player obtains $36$ pieces of six colors
and  of six different heights
and is supposed to place them in the square to obey 
all the rules of OLS - see Fig.~\ref{figcube36}.

\end{document}